\documentclass[acmlarge]{acmart}
\usepackage{amsmath}

\usepackage{enumitem}
\usepackage{gensymb}
\usepackage{subfig}
\usepackage{graphicx}
\usepackage{booktabs}
\usepackage{multirow}
\usepackage{makecell}
\usepackage{mathtools}
\DeclareMathOperator{\Prb}{Pr}
\usepackage[font=small,labelfont=bf]{caption}
\usepackage[per-mode=fraction]{siunitx}
\usepackage[labelfont=bf]{caption}
\AtBeginDocument{%
  \providecommand\BibTeX{{%
    \normalfont B\kern-0.5em{\scshape i\kern-0.25em b}\kern-0.8em\TeX}}}

\copyrightyear{2020}
\setcopyright{rightsretained}
\acmJournal{IMWUT}
\acmYear{2020} \acmVolume{4} \acmNumber{4}
\acmDOI{10.1145/3432234}

\begin{document}


\title{ComFeel: Productivity is a Matter of the Senses Too}


\author{Marios Constantinides}
\affiliation{%
  \institution{Nokia Bell Labs}
  \city{Cambridge}
  \country{UK}
}
\email{marios.constantinides@nokia-bell-labs.com}

\author{Sanja \v{S}\'{c}epanovi\'{c}}
\affiliation{%
  \institution{Nokia Bell Labs}
  \city{Cambridge}
  \country{UK}
}
\email{sanja.scepanovic@nokia-bell-labs.com}

\author{Daniele Quercia}
\affiliation{%
  \institution{Nokia Bell Labs}
  \city{Cambridge}
  \country{UK}
}
\email{daniele.quercia@nokia-bell-labs.com}

\author{Hongwei Li}
\affiliation{%
  \institution{Nokia Bell Labs}
  \city{Cambridge}
  \country{UK}
}
\email{hongwei.3.li@nokia-bell-labs.com}

\author{Ugo Sassi}
\affiliation{%
  \institution{Nokia Bell Labs}
  \city{Cambridge}
  \country{UK}
}
\email{ugo.sassi@nokia-bell-labs.com}

\author{Michael Eggleston}
\affiliation{%
  \institution{Nokia Bell Labs}
  \city{Murray Hill}
  \country{US}
}
\email{michael.eggleston@nokia-bell-labs.com}


\renewcommand{\shortauthors}{Constantinides et al.}

\begin{abstract}
Indoor environmental quality has been found to impact employees' productivity in the long run, yet it is unclear its meeting-level impact in the short term. We studied the relationship between sensorial pleasantness of a meeting’s room and the meeting’s productivity. By administering a 28-item questionnaire to 363 online participants, we indeed found that three factors captured 62\% of people's experience of meetings:  (a) \emph{productivity};  (b) \emph{psychological safety}; and (c) \emph{room pleasantness}. To measure room pleasantness, we developed and deployed ComFeel, an indoor environmental sensing infrastructure, which captures light, temperature, and gas resistance readings through miniaturized and unobtrusive devices we built and named `\emph{Geckos}’. Across 29 real-world meetings, using \emph{ComFeel}, we collected 1373 minutes of readings. For each of these meetings, we also collected whether each participant felt the meeting to have been productive, the setting to be psychologically safe, and the meeting room to be pleasant. As one expects, we found that, on average, the probability of a meeting being productive increased by 35\% for each standard deviation increase in the psychological safety participants experienced. Importantly, that probability increased by as much as 25\% for each increase in room pleasantness, confirming the significant short-term impact of the indoor environment on meetings' productivity.
\end{abstract}

\ccsdesc[500]{Human-centered computing~Ubiquitous and mobile devices}
\ccsdesc[500]{Human-centered computing~Computer supported cooperative work}
\ccsdesc[500]{Human-centered computing~Empirical studies in ubiquitous and mobile computing}
\ccsdesc[300]{Human-centered computing~User studies}
\ccsdesc[500]{Human-centered computing~Empirical studies in ubiquitous and mobile computing}
\ccsdesc[300]{Human-centered computing~User studies}

\keywords{productivity, meetings, environmental conditions, perceptions}

\maketitle

\section{Introduction}
\label{sec:introduction}
A prolonged exposure to poor indoor environmental conditions has been found to impact employees' cognitive functions~\cite{allen2016associations} and decision-making~\cite{macnaughton2015economic}, have a direct effect on workers' performance and productivity~\cite{fisk2000health,mujan2019influence}, and eventually lead to increased levels of absenteeism ~\cite{hantani2009effect,wyon2004effects,snow2018indoor}. Health-related problems such as asthma, headaches, throat irritation, respiration, and allergies~\cite{mitchell2007current,mendell2003environmental} have reportedly been associated with a phenomenon often referred to as the Sick Building Syndrome (SBS)~\cite{redlich1997sick, burge2004sick}. 

Traditionally, companies and organizations have resorted to sensors through which the environmental conditions could be sensed~\cite{alavi2017comfort, wyon2004effects, zhong2020hilo, hantani2009effect}, and even adapted accordingly~\cite{huizenga2006air, bader2019windowwall} to meet recommended standards\footnote{https://www.ashrae.org/technical-resources/bookstore/standards-62-1-62-2}, thus increasing their employees' productivity and well-being. Sensing devices are often being deployed in spaces to sense indoor environmental quality (IEQ). To circumvent shortcomings of commercial devices such as costs or flexibility of platforms in terms of data acquisition, IEQ devices based on open-source electronic designs were made available~\cite{karami2018continuous}. For example, \citet{ali2019elemental} proposed Elemental, an open-source device that combines IEQ readings with energy usage and HVAC operation sensors. A recent review of studies on indoor comfort~\cite{song2019human} identified the application of machine learning to IEQ factors other than thermal comfort as one of the research gaps. Our work partly fills this gap. Additionally, as we shall see in \S\ref{sec:related_work}, while previous studies highlighted the impact of the environmental conditions on people's productivity in the long run, it is unclear to which extent that impact translates in the short term at a meeting-level.

To explore the short term impact of indoor environmental quality, we looked at one of the most common daily activities at work: \emph{meetings}. More specifically,  we studied the relationship between the sensorial pleasantness of a meeting's room and the meeting's productivity. In so doing, we made four main contributions:

\begin{itemize}
    \item We operationalized the concept of `sensorial pleasantness' and `productivity' by developing an online 28-item questionnaire, grounded on literature in the fields of Management and Organizational Science, and administering it to 363 participants (\S\ref{sec:section3}). Based on factor analysis, we indeed found that three factors captured 62\% of people's experience of meetings: (a) \emph{productivity} (i.e., whether a meeting has been productive); (b) \emph{psychological safety} (i.e., whether participants felt listened to); and (c) \emph{room pleasantness} (i.e., whether the room was felt to be pleasing to the senses).
    \item We developed  an indoor environmental sensing infrastructure, called \emph{ComFeel}, which captures light, temperature, and gas resistance readings through miniaturized and unobtrusive devices that we built and named `Geckos'(\S\ref{sec:framework}). 
    \item To measure room pleasantness, we deployed ComFeel across 29 real-world meetings held at a corporate setting, and collected 1373 minutes of environmental readings (\S\ref{sec:user_study}). For each of these meeting, we also collected whether each participant felt the meeting to have been productive, the setting to be psychologically safe, and the meeting room to be pleasant.
    \item Using the collected data, we studied the interplay between the sensorial pleasantness and productivity (\S\ref{sec:analysis}, \S\ref{sec:results}). We found that, on average, the probability of a meeting being productive increased by 35\% for each standard deviation increase in the psychological safety participants experienced. Importantly, the very same probability increased by as much as 25\% for each standard deviation increase in room pleasantness. These results suggest that indoor environmental conditions do matter not only in the long run (as previous work found), but also in the short term so much so that significant differences in productivity were observed even within the constrained temporal space of individual meetings. 
\end{itemize}
\section{Related Work}
\label{sec:related_work}
Next, we surveyed various lines of research that our work draws upon, and we grouped them into three main areas: \emph{i)} indoor environmental sensing, \emph{ii)} the effects of IEQ on productivity and performance, and \emph{iii)} productivity and meetings.

\subsection{Indoor Environmental Sensing: Reality and Perception}
A prolonged exposure to poor indoor environmental conditions has been found to impact people's health~\cite{sundell2004history}, cognitive functions~\cite{allen2016associations} and  decision-making~\cite{macnaughton2015economic}, and eventually lead to increased levels of absenteeism ~\cite{hantani2009effect,wyon2004effects,snow2018indoor}. SBS is a term often used in Occupational Safety and Health\footnote{www.cdc.gov/niosh} to describe health-related problems associated with poor ``indoor air quality''. Symptoms such as watering eyes, headaches, dizziness, mental fogginess, respiration, and allergies, just to name a few, make up the SBS symptoms list. Report after report\footnote{https://buildingevidence.forhealth.org/} has shown that the amount of `good' air and, more broadly, good indoor environmental conditions can reduce SBS symptoms, cut absenteeism, and even put on hold the transmission of infectious diseases~\cite{hbr_office_air}. The most prominent indoor environmental properties identified in previous works include the ventilation rate, the particle concentration, the amount of carbon dioxide and volatile organic compounds, the temperature, the floor-coverings and used supply air filters, the computational equipment, the lighting conditions, and the noise distraction and occupancy rate~\cite{wyon2004effects,allen2016associations,macnaughton2015economic,sbar2012electrochromic, jin2016occupancy, jin2014environmental}. Links between the actual indoor environmental conditions and people's perceptions about it have also been reported in studies~\cite{de2014measured,kang2017impact}. For example, studies have shown differences between doctors' and patients' perceptions of the environmental conditions at the same hospital~\cite{de2013measured}, and gender differences when reporting perceived environmental quality (female employees consistently report lower levels~\cite{kim2013gender}).

\subsection{Long and Short-term Effects of IEQ on Productivity}

The effects of IEQ factors on productivity, performance, and absenteeism were studied at real-world workplaces~\cite{niemela2002effect,clements2000productivity,hantani2009effect,wyon2004effects,snow2018indoor}, or in controlled laboratory environments \cite{bako2004effects,wargocki2002ventilation,ishii2018intellectual}. \citet{neidell2017air} highlighted the importance of IEQ effects: \emph{`Although the damage per individual is small when compared to more extreme events, such as mortality and hospitalizations, the effects are more widespread and may thus represent an important cost to society.'} In addition to IEQ, another important concept is the perceived environmental quality (PEQ)~\cite{fanger1988introduction}: in some cases, even when employees could not consciously perceive poor IEQ conditions their productivity and performance were negatively impacted~\cite{wyon2004effects,seppanen2006some}. While the short-term effect of environmental quality on productivity in both experimental \cite{wyon2004effects} and real workplace \cite{niemela2002effect} settings was studied, it is yet unclear the short-term effect of indoor environments on a recurring activity within a constrained temporal space such as a \emph{work meeting}.

\subsection{Productivity and Meetings}
\label{subsec:lit_meetings}
In the workplace, performance is a multi-faceted construct mostly characterized by `how well workers and employees perform their tasks, the initiative they take and the resourcefulness they show in solving
problems'~\cite{viswesvaran2000perspectives, rothmann2003big}. Simply put, how productive they are. Traditionally, companies and organizations assess their workforce's productivity using various methods including self-reports, and peer- and supervisors-ratings~\cite{iaffaldano1985job, barrick1991big}. However, these methods are not always administered frequently, thus making it challenging for organizations to obtain a snapshot of their workforce's productivity in the short term, and are often debatable due to potential biases that self-assessments may entail~\cite{sartori2017not, fan2006exploratory}. More recently, it has been shown that job performance and productivity might well be assessed using passive data collected from workers' smartphones and wearable devices~\cite{mirjafari2019differentiating}, and workers' cognitive states can be successfully determined~\cite{schaule2018employing}. Generally speaking, measuring productivity, however, can be a complex, challenging task depending on the work activity one looks into. 

One of the most common daily activities at work is meetings. It is one of the most common practices in which organizations mobilize their individual members to work together, share ideas, brainstorm solutions, and solve problems. While people allocate significant amount of their work time in attending meetings, and organizations devote notable large amounts of resources to facilitate them~\cite{macnaughton2015economic}, people still tend to feel disconnected and perceive them as unproductive~\cite{romano2001meeting}. Reports suggest that one in every three employees, across cities and industry sectors, considers meetings ineffective and unproductive \cite{attentiv}. However, ineffective meetings are bad for morale and productivity, negatively impact employees' health~\cite{hbr_meeting_madness}, and are directly linked to organizations' wasted time and resources~\cite{kauffeld2012meetings, rogelberg2012wasted}. Therefore, it is paramount for organizations to understand how their employees experience meetings as an attempt to boost their productivity, and design tools to support it~\cite{meetcues}. 

Traditionally, a meeting's performance and perceived experience has been evaluated through standardized questionnaires, or ad-hoc post meeting surveys. For example, the Meeting Effectiveness Index (MEI)\footnote{MEI: http://www.meetingmetrics.com/EPI-Report-demo-data.pdf} is a six-item questionnaire that focuses on a meeting's overall performance results. In the fields of Management and Organizational Science, meetings' productivity has been assessed by probing its agenda, structure, and purpose~\cite{lent2015_meetings, hbr_cohen_start, hbr_schwarz_start}.
Inclusiveness~\cite{hbr_better_meetings, hbr_quality_experience}, dominance~\cite{romano2001meeting}, peripheral activities~\cite{niemantsverdriet2017recurring}, and psychological safety~\cite{hbr_tense, hbr_psychological_safety} are aspects that have also been found to influence meetings' experience and productivity. The physical comfort of the workplace environment is an additional one that makes up the list~\cite{alavi2017comfort}. For example, a recent survey\footnote{https://futureworkplace.com/} found that light and outdoor views were among the most popular perks employees craved for~\cite{hbr_office_light}. Furthermore, studies have shown that stuffy and stale air offices reduce employees' productivity~\cite{allen2016associations, hbr_office_air, hbr_air_pollution}.

\section{Online Survey}
\label{sec:section3}

\begin{table}
    \captionsetup{labelfont=normalfont}
	\begin{minipage}{0.5\linewidth}
		\caption{Demographics of the Online Survey Participants (\emph{Note: Five participants did not disclose their age.)}}
\label{tbl:survey_demographics}
		\centering
		\begin{tabular}{lll}
    \toprule
    \textbf{} & \textbf{Range} & \textbf{\#answers}  \\
    \midrule
    \multirow{4}{*}{Age} & 18-24 years old & 3 \\
    & 25-34 years old  & 122\\
    & 35-44 years old & 84\\
    & 45-54 years old & 50\\
    & 55-64 years old & 19\\
    & 65+ years old & 5\\
		\midrule
    \multirow{3}{*}{Gender} & Female  & 154\\
    & Male & 132\\
    & Prefer not to say & 2\\
		\midrule
    \multirow{2}{*}{Education} & High school graduate & 45 \\
    & Bachelor's degree &  151 \\
    & Master's degree &  63 \\
    & Ph.D. or medical degree &  29 \\
  \bottomrule
\end{tabular}
	\end{minipage}\hfill
	\begin{minipage}{0.45\linewidth}
	    \captionsetup{labelfont=normalfont}
		\centering
		\includegraphics[width=7cm,keepaspectratio]{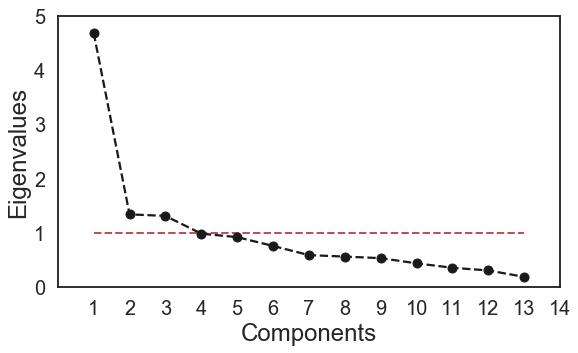}
    \captionof{figure}{Scree plot showing our PCA components and their eigenvalues in descending order.}
	\label{fig:scree_plot_leveling}
	\end{minipage}
\end{table}

As stated in the previous section, existing questionnaires used to evaluate meetings focus primarily on performance, whereas the perceived experience of attendees has received little attention. We reviewed previous works from HCI communities, and articles in the fields of Management and Organizational Science relating to `meetings'~\cite{hbr_better_meetings, hbr_quality_experience,romano2001meeting, niemantsverdriet2017recurring, hbr_tense,hbr_psychological_safety, hbr_air_pollution, hbr_office_air, hbr_office_light}, and we identified themes concerning people such as their psychological experience (i.e., hierarchy and cultural balance, personality and gender diversity, and contribution and broad/participation), and the meetings' structure and content (i.e., agenda, environment, technology used, action points, time, and multitasking). Based on this review, we developed a questionnaire\footnote{http://bit.ly/psycho-meetings} consisting of 28 questions that cover the wide range of themes we identified, and an additional set of three questions relating to demographics (age, gender, and education levels). 
\subsection{Participants and Recruitment}
We distributed the survey via social media (i.e., Facebook and Twitter on targeted lists of knowledge workers. We specified `job meeting' and `project management' as interests, and targeted a range of sectors from administrative to sale to legal services to business and finance), and as an Amazon Mechanical Turk (AMT) task. We ensured highly reputable AMT workers by targeting workers with 95\% HIT approval rate, and at least 100 approved HITs. We received a total of 363 responses from both sources. We applied quality checks in the form of trap questions. For example, one question prompted participants to report whether they attended the meeting remotely or in-person, while a subsequent question asked them to state their perceived air quality of the room where the meeting took place. We rejected those who reported contradicting answers (e.g., attended the meeting remotely and answered the air quality question, while having the option to skip the air quality question), thus ensuring data quality. As the survey questions were not mandatory, we carried out further filtering of empty responses. We also eliminated those whose last meeting was held `more than a week ago' to remove potential recall bias. The final dataset we analyzed consisted of 288 responses (Table~\ref{tbl:survey_demographics}).

\begin{table*}
\captionsetup{labelfont=normalfont}
\caption{Rotated Component Matrix for the 3 derived principal components of the psychological experience during meetings: \newline \emph{Psychological safety}; \emph{Productivity}; and \emph{Pleasantness}.}
\centering
  \resizebox{\textwidth}{!}{
  \begin{tabular}{llll}
    \toprule
    \textbf{} &\vtop{\hbox{\strut \textbf{Psychological safety}}}& \textbf{Productivity} & \textbf{Pleasantness}\\
    \midrule
   \makecell[l]{1. I received the appropriate attention when \\ sharing my thoughts.}  & .835   &  &    \\
   \makecell[l]{2. I felt comfortable sharing my thoughts \\ and making contributions.}  & .814   &  &    \\
   \makecell[l]{3. I felt that my peers were comfortable sharing \\ their thoughts and making contributions.}  & .804  &  &    \\
   4. I found the meeting to be useful for all attendees.  & .774  &  &    \\
   5. I found the meeting to be useful for me.  & .747  &  &    \\
   6. I felt overall satisfied during the meeting.  & .528  &  &    \\
   &&& \\
   7. Summary points were sent out before the meeting ended.   &   & .773  &    \\
   8. Summary points were sent out after the meeting has ended.  &   & .647 &    \\
   9. The meeting had a clear purpose. &   & .611  &    \\
   &&&\\
   10. Amount of space in the meeting room. &   &   &   .865  \\
   11. Air quality in the meeting room. &   &   &   .790  \\
   \midrule
   \emph{component loadings below .5 are suppressed.} & & & \\
  \bottomrule
\end{tabular}}
\label{tbl:rotated_component_matrix}
\end{table*}

\subsection{Principal Component Analysis}
We performed a Principal Component Analysis (PCA) on the survey responses to identify the main orthogonal dimensions that explain the responses. PCA ``determine[s] the linear combinations of the measured variables that retain as much information from the original measured variables as possible.''~\cite{fabrigar1999evaluating}. The filtered dataset contained 288 responses and a ratio of 10.2 cases per variable, well above the recommended ratio of 1:10 for PCA analysis~\cite{nunally1978psychometric}.

To ascertain the validity of our PCA results, we conducted three tests. First, we tested the correlation matrix between the 28 questions of the survey (excluding demographics). It showed significant correlations of above .3 with at least one other item for the 20 out of 28, suggesting reasonable factorability. Secondly, we used Bartlett's Test of Sphericity which was found to be significant (${\chi}^2$(378) = 3092.43, p = .000). Thirdly, the Kaiser-Meyer-Olkin measure of sampling adequacy (MSA) was .716, which was above the recommended threshold of .7~\cite{lloret2017exploratory}. Subsequently, fifteen items with MSA below .7 in the anti-image correlation matrix were excluded from the analysis~\cite{norman2008biostatistics}. These items included questions related to a respondent's current position (e.g., years of employment, employer); and the meeting's diversity (e.g., number of attendees, gender), interactions/conversation style, and time of day. This exclusion resulted in improved values of (${\chi}^2$(78) = 1302.54, p = .000) in Barlett's Test of Sphericity, and in an overall MSA score of .859. This made the PCA more robust with the remaining 13 items.

To determine the number of components that explain the most variance in our survey, we inspected the eigenvalues of the principal components (Figure~\ref{fig:scree_plot_leveling}). The eigenvalue of a principal component reflects the variance in all the variables explained by this component, and a higher value indicates higher variance in the variables loaded together in this component. The Kaiser's rule~\cite{kaiser1960application} suggests eigenvalues above 1.0 to be retained (resulting in 4 principal components), while the Cattell~\cite{cattell1966scree} proposed a visual inspection of the Scree Plot (Figure~\ref{fig:scree_plot_leveling}) to determine the `elbow' point in which a leveling effect is observed. This resulted in 3 principal components that explained 56.12\% of the total variance in the data. We considered the loadings of the 13 questionnaire items of the 3 principal components. The loadings describe the correlations between a single item and the principal component. We removed 2 items from further analysis due to their low component loading below .5. This resulted in 11 remaining items, which accounted for 62.31\% of the total variance; a solution of 60\% variance explainability is often considered as satisfactory in social sciences (\cite{hair1998multivariate} p.107). The additional two items left out concerned the respondent's meeting role (e.g., organizer, attendee), and the level of excitement during their last meeting, either of their loading components was below .5.

Table~\ref{tbl:rotated_component_matrix} shows the resulting components and their items. The first component is about \emph{psychological safety}, which refers to personal and group's safety of contributing to the meeting, personal and groups' views on usefulness, attention received, and personal overall satisfaction. The second component is about \emph{productivity}, which refers to whether the meeting had clear purpose and structure, and resulted in a list of actionable items (i.e., whether a meeting has been productive). The third component is about \emph{pleasantness}, which refers to the meeting's room overall pleasantness (e.g., good indoor air quality and physical space utilization).

\subsection{Three Factors of Meetings' Psychological Experience}
\label{subsec:factors}
To make use of these three PCA components in our user study (\S\ref{sec:user_study}), we paraphrased them into three questions our participants are likely to understand without considerable effort:
\begin{itemize}[leftmargin=9mm, label={}]
    \item $Q_{psychological}$: Did you feel \emph{listened to} during the meeting or \emph{motivated to be involved in} it?
    \item $Q_{productive}$: Did the meeting have a \emph{clear purpose and structure}, and did it result in a list of \emph{actionable points}?
    \item $Q_{pleasantness}$: Did the \emph{room feel pleasant} (in regards to air quality and crowdedness)?
\end{itemize}
\section{ComFeel Framework}
\label{sec:framework}
\begin{figure*}
\captionsetup{labelfont=normalfont}
\centerline{\includegraphics[width=0.7\linewidth]{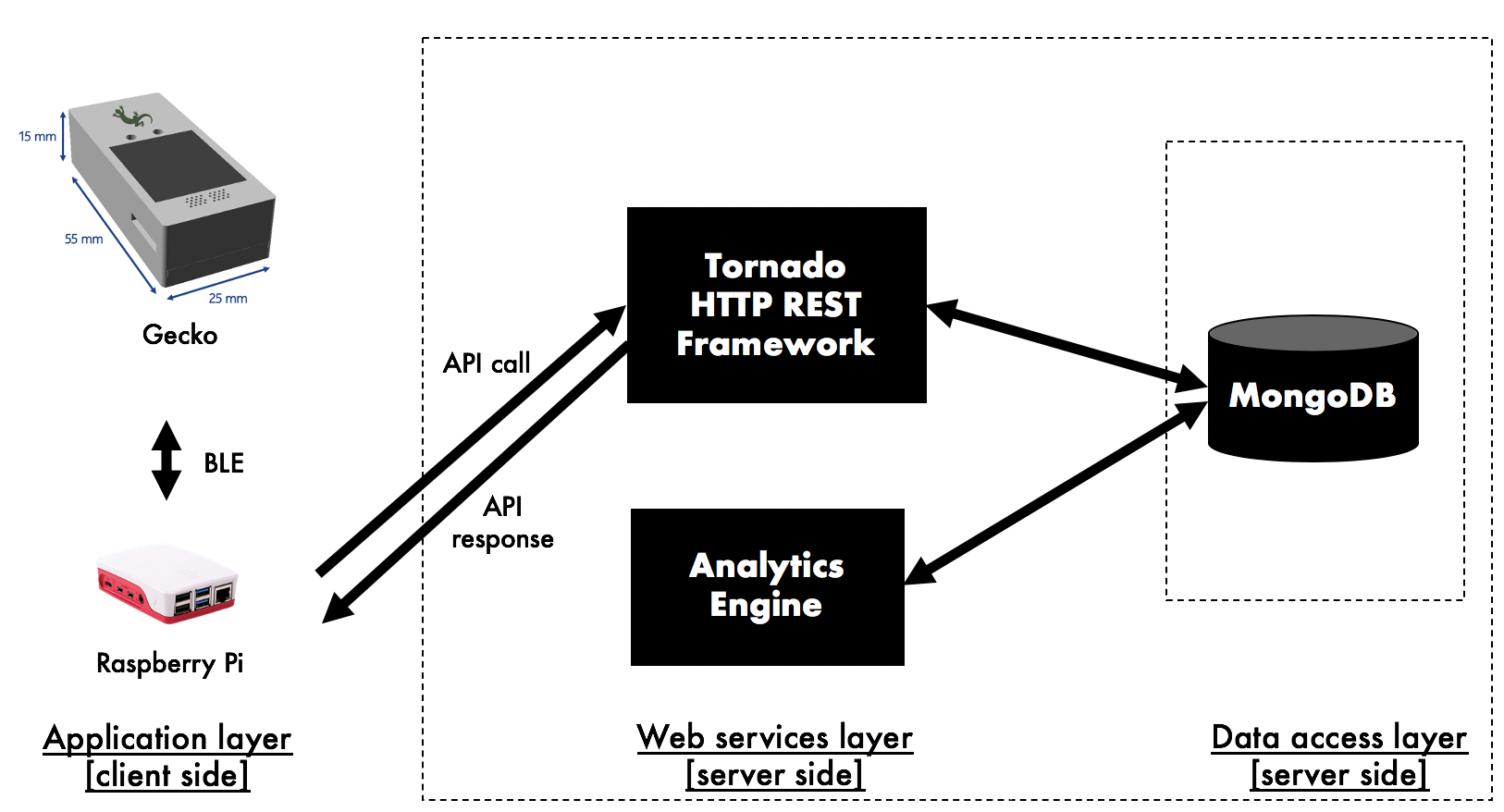}}
\caption{ComFeel sensing infrastructure. Our Gecko device continuously monitors indoor environmental properties (i.e., light, temperature, gas resistance), which are transmitted to an intermediate agent implemented using a Raspberry Pi. Upon data acquisition, the Raspberry Pi transmits the data to our backend infrastructure for further processing.}
\label{fig:sensing_framework}
\end{figure*}

As our study aims to investigate the short-term effects of sensory layers on people's productivity in the context of meetings, we resorted to sensors to obtain objective measurements about the indoor environmental conditions. In what follows, we describe our data collection framework, namely ComFeel. It is a three-tier framework (Figure~\ref{fig:sensing_framework}) composed by an application layer (a Gecko device and a Raspberry Pi), a web service layer, and a data access layer. 

\subsection{Application Layer}
\subsubsection{Gecko Device}
\label{subsec:gecko}
We assembled Gecko, a small (55x25x15), wearable environmental device, using off-the-shelf commercial sensors. By building our own device with low-manufacturing costs, we had full control of the firmware and we owned the raw data, which is often not the case in commercial devices. Additionally, its flexible design and plug\&play functionality make it easy to deploy, and attractive for large-scale deployments. Its total costs is 88 USD, which includes a solar cell (2 USD), a battery (2 USD), a plastic case (8 USD), and a PCB board with all the electronic components (76 USD). This estimation is based on production of 50 units, whereas, in case of mass production (\char`\~1000 units), the costs could further be reduced to 50 USD. Assembly costs are not included in the aforementioned costs, as they are deemed negligible due to a Gecko's design being assemblable in less than 10 minutes.

We used Bosch's environmental sensor BME680\footnote{https://www.bosch-sensortec.com/products/environmental-sensors/gas-sensors-bme680/} for monitoring indoor temperature levels (C\degree) and gas resistance (ohm), and a BH1745NUC digital 16-bit serial output type color sensor IC for obtaining light luminosity\footnote{Color sensor: https://www.mouser.co.uk/datasheet/2/348/bh1745nuc-e-519994.pdf} (fc: foot-candle). We configured both sensors to sample every 60 seconds, and sensors readings were obtained through a micro-controller using an I2C interface on the predefined interval. Upon reading, Gecko packs up each sensor's raw readings and sends it over using a Bluetooth Low Energy (BLE) protocol. Given that the payload of each BLE packet cannot exceed 244 bytes, as specified in the Bluetooth standard 4.2 onwards, each sensor reading is wrapped up in a packet. Therefore, a packet contains up to 244 bytes of information, and has the following signature: [packet identifier, timestamp, sensor id, sensor data length], where \emph{packet identifier} indicates that a packet carries sensor data payload, the \emph{timestamp} contains the time in which the reading occurred, the \emph{sensor id} is a unique identifier for each sensor, and the \emph{sensor data length} is the actual reading in bytes. When the packet is ready, we used a Nordic UART service\footnote{https://developer.nordicsemi.com/nRF\_Connect\_SDK/doc/latest/nrf/include/bluetooth/services/nus.html} to transmit it to an intermediate agent (i.e., Raspberry Pi), which we describe next. The choice of the Nordic UART service was primarily reinforced by the fact that it is an easy to implement and adapt it to our streaming requirements, which are independent of the type of reading and the sampling rate. Additionally, the BLE protocol itself ensures that packet loss or data errors during transmission would not happen. At the end of every link layer packet there is a 24-bit CRC; a mechanism that BLE uses, which contains a sequence number and the next expected one. Every packet is then acknowledged at the link layer, thus if no acknowledgment is received the packet will be retransmitted. To further validate any potential errors due to packet drops, in a subsequent section (\S\ref{subsubsec:rasppi}), we describe the reliability of the Raspberry Pi gateway.

\begin{figure*}
    \captionsetup{labelfont=normalfont}
    \centering
    \subfloat[]{
        \includegraphics[width=0.32\textwidth ]{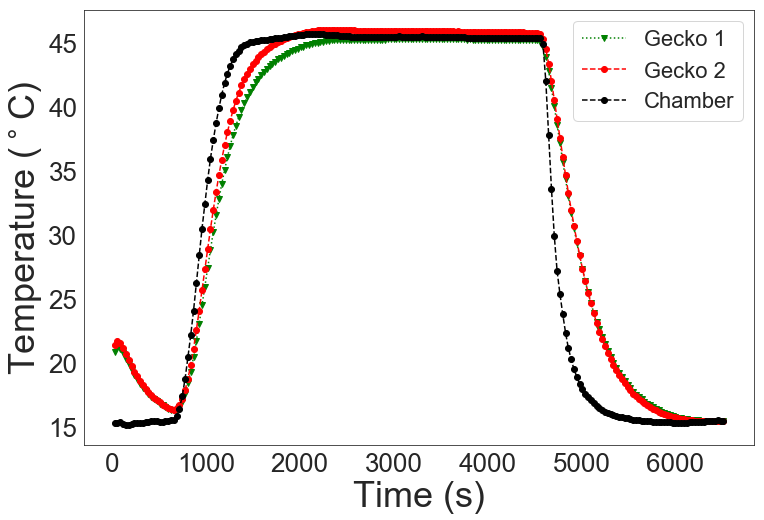}
        \label{fig:temp0}
    }
    \subfloat[]{
        \includegraphics[width=0.32\textwidth ]{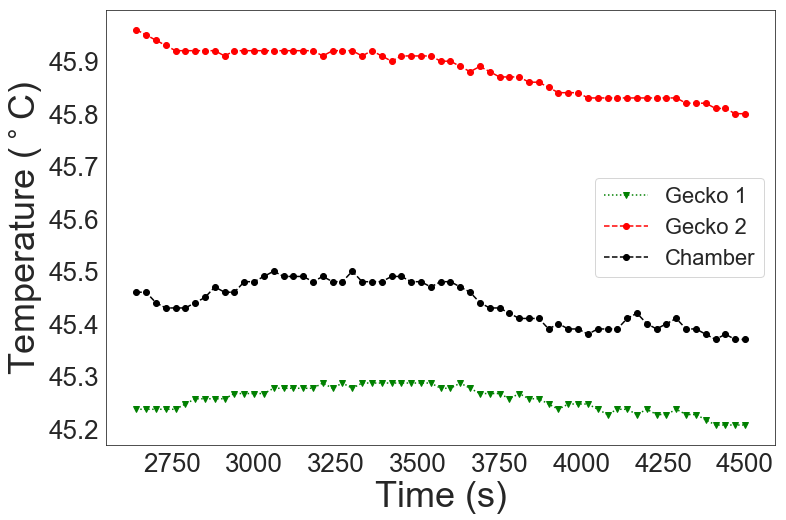}
        \label{fig:temp1}
    }
    \subfloat[]{
        \includegraphics[width=0.32\textwidth ]{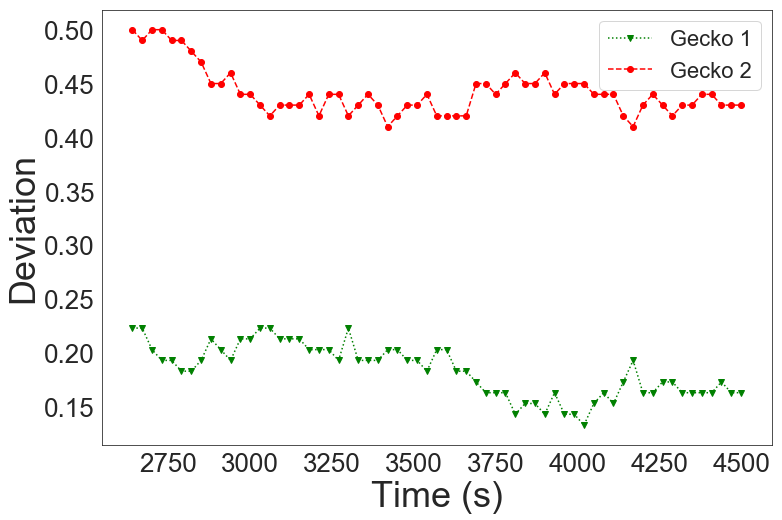}
        \label{fig:temp2}
    }
    \caption{(a) Geckos' vs. environmental chamber temperature measurements at 15\celsius ~and 45\celsius, (b) zoomed-in temperature measurements at 45\celsius ~for which the accuracy was estimated, and (c) deviations of the two Gecko devices against the environmental chamber's measurements.}
    \label{fig:temp_experiment}
\end{figure*}

\subsubsection{Gecko Sensors Reliability}
To ascertain our device's sensors reliability, we carried out experiments that compared our devices' readings against the measurements operated in an environmental chamber; a typical instrument for reliability testing of different devices at different temperatures~\cite{lee2001characterization}. The accuracy of the temperature sensor inside the chamber is of 0.1\celsius.

According to the manufacturer\footnote{https://www.bosch-sensortec.com/products/environmental-sensors/gas-sensors-bme680/}, the temperature sensor of our Gecko has an operating range from -40\celsius ~to 85\celsius ~with an accuracy of 1\celsius; the gas sensor has a response time of 1s, and an accuracy of 0.08\%; and the light sensor is capable to measure a wide dynamic lux range up to 40K lux with an accuracy of 0.005\footnote{https://www.mouser.co.uk/datasheet/2/348/bh1745nuc-e-519994.pdf}.

To test the temperature sensor readings, we placed two Gecko devices in the environmental chamber. We set the chamber's temperature to 15\celsius ~and waited until it stabilized in order to place the devices inside. After 10 minutes of stabilization, we increased the temperature to 45\celsius ~and, in turn, after roughly 50 minutes, we decreased it again to 15\celsius ~for another 10 minutes (Figure~\ref{fig:temp_experiment}a). Using the values obtained at 15\celsius, ~we calibrated our sensors, and computed the accuracy of our temperature sensor from the measurements at 45\celsius ~(Figure~\ref{fig:temp_experiment}b). We then estimated the accuracy as the mean value of all deviations between the Gecko measurements and the chamber temperature values at 45\celsius~ (Figure~\ref{fig:temp_experiment}c). We observed worst-case precision of 0.5\celsius, which is in the range of the accuracy stated by the manufacturer\footnote{https://www.bosch-sensortec.com/products/environmental-sensors/gas-sensors-bme680/}. Additionally, we observed that the two Gecko devices showed very similar response, which suggests low sensor-to-sensor variability.

Taking into account typical illumination levels of office spaces (i.e., in the range of 500 lux~\cite{mui2006acceptable}), and given the capability of our light sensor to measure a wide dynamic lux range from 0.005 to 40K lux with an accuracy of 0.005 lux\footnote{https://www.mouser.co.uk/datasheet/2/348/bh1745nuc-e-519994.pdf}, even an error of one order of magnitude would still provide accurate measurements. The same applies to the high accuracy of the gas resistance measurement as stated by the manufacturer (i.e., 0.08\%\footnote{https://www.bosch-sensortec.com/products/environmental-sensors/gas-sensors-bme680/}).

\subsubsection{Raspberry Pi}
\label{subsubsec:rasppi}
As Gecko is designed to be a cheap and flexible indoor environmental device, it has no capabilities of streaming its data to a web service. To that end, we implemented an intermediate agent using Raspberry Pi which acts as a gateway between a Gecko device and our backend server. The Raspberry Pi collects data from one or more Gecko devices and forwards it to a web server. The gateway is implemented in such a way to actively listen for connections over BLE. Once a Gecko device establishes a BLE connection with the Raspberry Pi, the gateway decodes the bytes array sent from Gecko, formats it in as a JSON, and transmits it to our backend for further processing and analysis. To further validate whether any potential errors occurred at the Rasperry Pi side, we transmitted data to the gateway from two Gecko devices. (Figure~\ref{fig:sensing_framework}). In total, both devices transmitted 1178 packets, and we observed 5 misses out of 587 packets sent from the first device, and 8 misses out of 591 packets sent from the second device. On average, this accounts for no more than 1.1\% error rate at the gateway level.

\subsection{Web Services Layer}
Using the Python Tornado framework, we developed a RESTful API that runs continuously on a server. It exposes an endpoint that handles the communication with the Raspberry Pi module and the data access layer. The endpoint is responsible for receiving the data from the gateway and flushing it to the database.

\subsection{Data Access Layer}
The data acceess layer is implemented by a MongoDB instance, and the raw data is stored in a database collection with the following generic schema: [deviceID, sensor\_type, sensor\_value, ts], where \emph{deviceID} is a unique identifier that distinguishes Gecko devices, \emph{sensor\_type} is a key identifier that corresponds to the type of environmental data (i.e., temperature, gas resistance, or light luminosity),  \emph{sensor\_value} is the sensor reading, and \emph{ts} is the time that the sensor reading was sampled. 

\begin{figure*}
    \captionsetup{labelfont=normalfont}
    \centering
    \subfloat[]{
        \includegraphics[width=0.4\textwidth ]{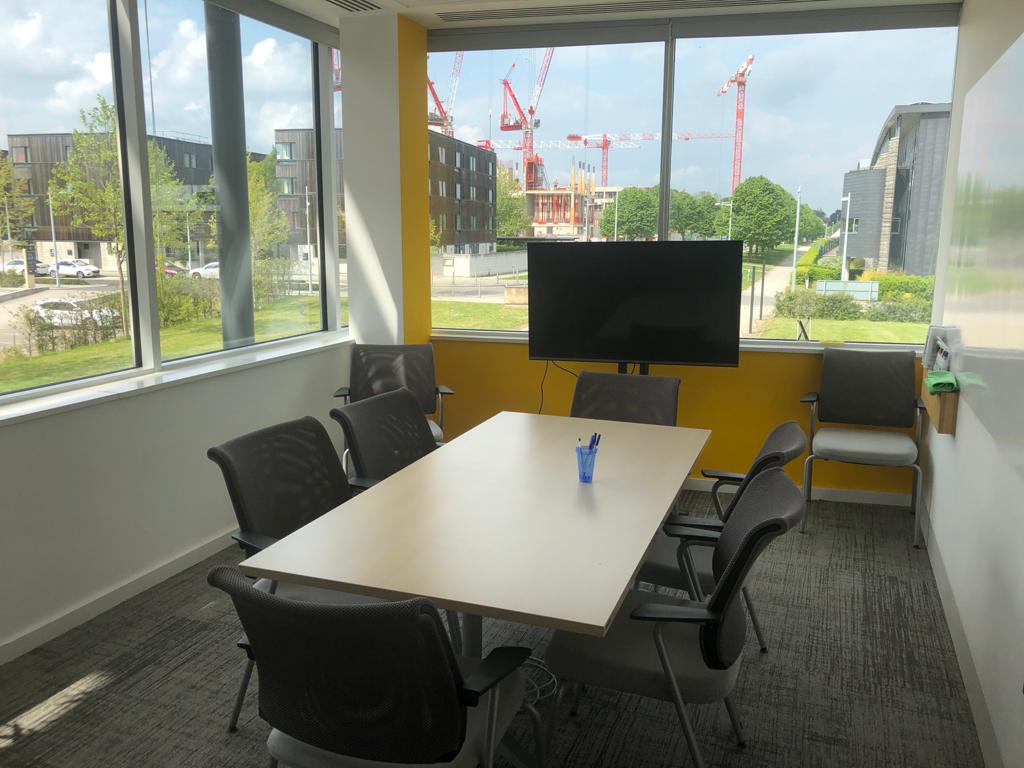}
        \label{fig:ada}
    }
    \subfloat[]{
        \includegraphics[width=0.4\textwidth ]{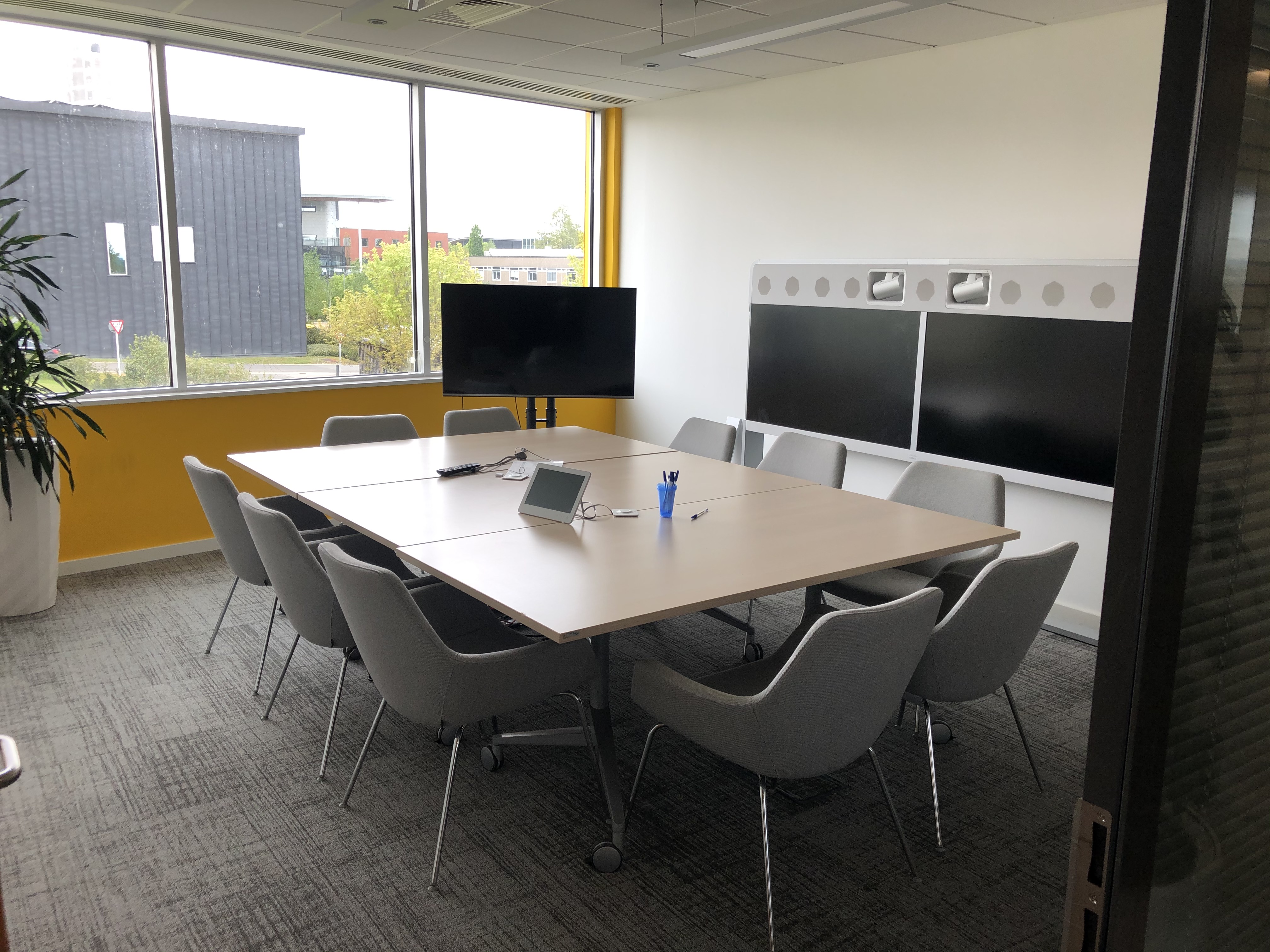}
        \label{fig:jocelyn}
    }
    \caption{The two meetings rooms utilized in our user studies. (a) the \emph{small} room has a maximum capacity of six people (\SI{16.25}{\metre\squared}), while (b) the \emph{large} room has a maximum capacity of ten people (\SI{24.70}{\metre\squared}).}
    \label{fig:rooms}
\end{figure*}

\section{User study}
\label{sec:user_study}
Having developed a sensing infrastructure to obtain objective measurements of indoor environmental quality, we conducted a longitudinal study, spanned over a period of one month, to understand the sensorial pleasantness of a meeting's room and the meeting's productivity.

\subsection{Study Setup}
\label{subsec:setup}

The study was conducted at the facilities of a research laboratory in a corporate setting during February-March. Two meeting rooms with varying capacities and dynamics were utilized (Figure~\ref{fig:rooms}). We refer to them as the \emph{small} room and the \emph{large} room. The small room's is \SI{16.25}{\metre\squared}, which is ideal for small-size meetings (six seats), while the large room's ones is \SI{24.70}{\metre\squared} (ten seats), which is more preferable for hosting larger-size meetings. The choice of these two particular rooms sizes was reinforced by their popularity. Therefore, we did not set up any `experimental' rooms, rather we deployed our framework in rooms people regularly use. This also allowed us to understand people's behavior and perception under real-world corporate settings as opposed to controlled lab settings. Additionally, these two rooms, when empty, are similar in terms of light (same view), temperature, and gas resistance. We also kept both rooms' blinds open during the trial, thus ensuring similar exposure to outside views in both rooms. We deployed two Gecko and Raspberry Pi devices, one in each room, which were continuously monitoring and storing the indoor environmental information on our backend server (Figure~\ref{fig:sensing_framework}).

\begin{table*}
\captionsetup{labelfont=normalfont}
\caption{The four-item survey which our participants completed at the end of each meeting.}
\label{tbl:questionnaire}
  \begin{tabular}{lll}
    \toprule
    \textbf{Name} & 
    \textbf{Question} & 
    \textbf{Responses} 
    \\
    \midrule
     $Q_{type}$ & What was the type of the meeting you just had? &
     {\shortstack[l]{ (a) status update \\ (b) information sharing /presentation \\ (c) decision making/problem solving }}\\  
     \midrule
     $Q_{productive}$ & {\shortstack[l]{ Did the meeting have a clear purpose and structure, \\ and did it result in a list of actionable points? }}   &
     {\shortstack[l]{ Likert Scale \\ (1: Strongly disagree; 7: Strongly agree)}} \\ 
     \midrule
     $Q_{psychological}$
     &
     {\shortstack[l]{ Did you feel \emph{listened to} during the meeting, or \\\emph{motivated to be involved in} it? }}   &
     {\shortstack[l]{ Likert Scale \\ (1: Strongly disagree; 7: Strongly agree)}} \\ 
     \midrule
     $Q_{pleasantness}$ & {\shortstack[l]{ Did the room feel pleasant?\\(in regard to air quality and crowdedness) }}   &
     {\shortstack[l]{ Likert Scale \\ (1: Strongly disagree; 7: Strongly agree)}}\\  
  \bottomrule
\end{tabular}
\end{table*}

\subsection{Materials and Apparatus}
To capture our participants' perceptions during a meeting, we asked them the three questions we identified in the online survey (\S\ref{sec:section3}), at the end of their meeting. We included an additional question to report the type of the meeting. All questions along with their response options are listed in Table~\ref{tbl:questionnaire}. Finally, participants noted in the survey the start and end time of their meeting, which we made use in our analysis (\S\ref{sec:analysis}). Participation in the questionnaires was anonymous. We opted-in for completely anonymized responses to ensure privacy compliance according to the EU General Data Protection Regulation (GDPR). As we did not have any particular requirements for participation, we neither adopted a recruitment protocol nor specified any eligibility criteria; instead, we simply installed an information sheet in both rooms, detailing the study's protocol.

\subsection{Protocol, Procedure, and Participants}
The protocol stated the aims of the study, the data collection, and made explicit to participants that they agree that their responses will be analyzed for research purposes. In accordance to GDPR, no researcher involved in the study could have tracked their identity by any means, and all their anonymous responses were analyzed at an aggregated level. In addition to the data handling and data analysis, the protocol stated the experimental procedure. Participants were informed that, for a period of one month, two Gecko devices were installed in the two meeting rooms, and these devices were monitoring both rooms' environmental conditions. In total, we received 364 self-reported answers to the four questions (Table~\ref{tbl:questionnaire}) from 91 participants who attended those meetings (Table~\ref{tbl:meetings_stats}).

\subsection{Dataset}
In total, we collected 1373 minutes of environmental data across 29 unique meetings that were held in the two meeting rooms (Table~\ref{tbl:meetings_stats}). As there is no unified classification of meeting types, we grouped our meetings into three common meeting types found in previous literature~\cite{meeting_sift, meeting_lifesize, meeting_thinkgrowth, meeting_minute}. These are: (a) \emph{status update}, (b) \emph{information sharing/presentation}, and (c) \emph{decision making/problem solving}. In total, our participants attended 58.2\% status update (\emph{type 0}), 28.6\% information sharing/presentation (\emph{type 1}), and 13.2\% decision making/problem solving (\emph{type 2}) meetings (Figure~\ref{fig:freq_dist}a), with varying duration (Figure~\ref{fig:freq_dist}e). To assess the quality of the responses to our three questions, we looked at the inter-rater agreement (IRA) among participants of the same meeting. We computed an IRA statistic called $r_{wg}$ score \cite{james1984estimating} and the standard deviation across participants' answers, which are commonly applied IRA measures \cite{o2017overview} for Likert-scale questionnaires. The $r_{wg}$ statistic (\ref{eq:rwq}) is a function of the observed variance in participants' answers ($S_x^2$), and the variance in the ratings, if they were random, i.e., null distribution ($\sigma_{eu}^2$):
\begin{gather}
\label{eq:rwq}
  r_{wg} = 1 - (S_x^2 / \sigma_{eu}^2), 
\end{gather} 
where $x$ is the set of the participants' answers, $S_x^2 = var(x)$, and $\sigma_{eu}^2$ is set by convention to a constant value of $(A^2 -1)/12$, and $A$ is the number of Likert-scale options (in our case, $A=7$). The values of $r_{wg}$ range from 0 to 1, with those closer to 1 representing a better agreement. The distributions of scores across meetings shown in Figure \ref{fig:ira} reveal a relatively high level of agreement among our participants for all the three questions (average $r_{wg}$ for $Q_{psychological}$ is .78, for $Q_{productive}$ is .87, and for $Q_{pleasantness}$ is .75). Unsurprisingly, our participants agreed more on the meeting's productivity than on the other two more subjective aspects.

\begin{table}
\captionsetup{labelfont=normalfont}
\caption{Statistics of the twenty nine meetings held in the two meeting rooms.}
\label{tbl:meetings_stats}
  \centering
  \begin{tabular}{lcc}
    \toprule
    & \textbf{small room} & \textbf{large room}\\
    \hline
    \#Meetings & 18 & 11\\
    \#Minutes of sensing data & 843 & 530 \\
    \#responses & 220 & 144\\
    \#participants responded & 55 & 36 \\
  \bottomrule
\end{tabular}
\end{table}

\section{Analysis}
\label{sec:analysis}

\begin{figure*}
    \captionsetup{labelfont=normalfont}
    \centering
    \subfloat[]{
        \includegraphics[width=0.25\textwidth ]{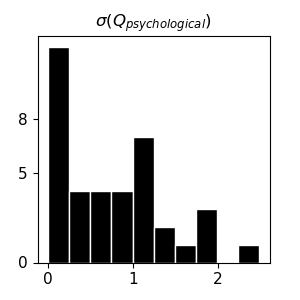}
        \includegraphics[width=0.25\textwidth ]{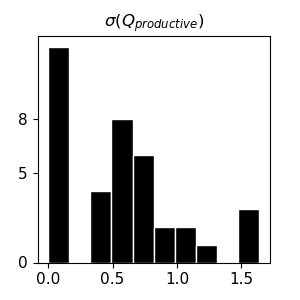}
        \includegraphics[width=0.25\textwidth ]{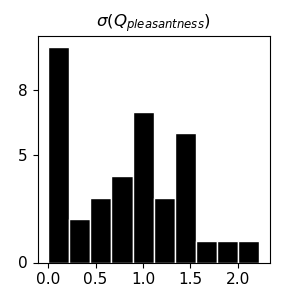}
        \label{fig:ira_stdev}
        }
        
        \subfloat[]{
        \includegraphics[width=0.25\textwidth ]{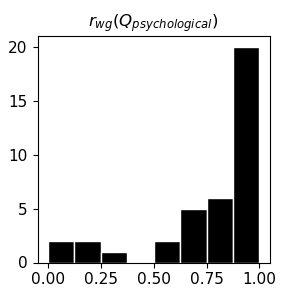}
        \includegraphics[width=0.25\textwidth ]{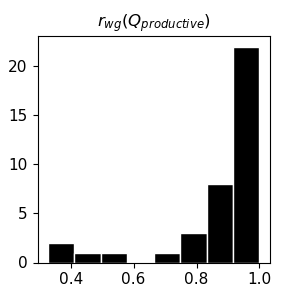}
        \includegraphics[width=0.25\textwidth ]{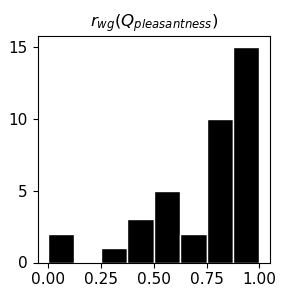}
        \label{fig:ira}
    }
    \caption{Distributions of inter-rater agreement among meeting participants for the three questions in terms of: (a) standard deviation $\sigma$ of their answer scores, and (b) the Likert-scale inter-rater agreement measure $r_{wg}$ \cite{james1984estimating}.}
    \label{fig:ira}
\end{figure*}

Having collected objective environmental measurements and employees' self-reported experiences of their meetings in terms of their perceived psychological safety, productivity, and room pleasantness, we set out to investigate the interplay of these factors with productivity. Based on prior work, we formulated three hypotheses, and defined a set of metrics to test them. 

\subsection{Hypotheses and Metrics}
\label{subsec:hypotheses}
As one expects, the outcome of a meeting (i.e., productive or not) is dependent on its type. In the fields of the Management and Organizational Science, the most common types include \emph{status update}, \emph{information sharing/presentation}, and \emph{decision making/problem solving}~\cite{meeting_sift, meeting_lifesize, meeting_thinkgrowth, meeting_minute}. It is also known that the length of a meeting might influence its productivity. Often, meetings that do not start or end on time tend to leave attendees' unsatisfied~\cite{hbr_cohen_start, hbr_schwarz_start}. Therefore, we formulated our first hypothesis as:\newline\newline\noindent\textbf{H\textsubscript{1}}:  Whether a meeting is productive depends on the meeting's \emph{type} and \emph{duration}.

To test it, we used a meeting's \emph{type} by asking our participants the $Q_{type}$ in Table~\ref{tbl:questionnaire}, and \emph{duration} computed from their timestamped responses for each meeting (Figure~\ref{fig:freq_dist}e). Since the \emph{meeting type} question (Figure~\ref{fig:freq_dist}a) is a categorical variable, we converted it into three dummy variables to then make use of them in a logistic regression: \emph{type 0}, \emph{type 1}, and \emph{type 2}.

We treated this problem as a classification task, and we used the binarized self-reported $Q_{productive}$ as the dependent variable. To binarize it into a positive and negative class, we defined a set $E$ as the set of all participants' responses to $Q_{productive}$, and we partitioned it into two groups such that $E = E^{>} \cup E^{<}$, where $E^{>}$ is the set of participants who reported higher levels of meetings' productivity (above the median) defined as $E^{>} = \{{e \mid > \bar{e}\}}$, where $e$ is a participant's $Q_{productive}$ response, and $E^{<}$ is the set of participants who reported lower levels of of meetings' productivity (below the median) defined as $E^{<} = \{{e \mid > \bar{e}\}}$, where $e$ is a participant's $Q_{productive}$ response. Then, the $E^{>}$ responses were assigned to the positive class, while the $E^{<}$ ones to the negative (Figure~\ref{fig:freq_dist}c).

We built the $M_{1}$ logistic regression model (\ref{eq:M1}) to test $H_{1}$ by including meetings' \emph{duration} and the three dummy variables (i.e., \emph{type 0}, \emph{type 1}, and \emph{type 2}), which reflect the \emph{type} of meeting:

\begin{gather}
  \shortintertext{\quad$ \Prb(\text{Q\textsubscript{productive)}} = 1 \mid \text{type 0, type 1, type 2, duration}) = $} \frac{1}{1 + exp^-\textsuperscript{($\alpha$ \text{ + } $\beta_{1}$ $type 0$ \text{ + } $\beta_{2}$ $type 1$ \text{ + } $\beta_{3}$ $type2$ \text{ + } 
$\beta_{4}$ $duration$ )}}, \label{eq:M1}
\end{gather}
where Pr is the posterior probability of an instance belonging to the positive class of $Q_{productive}$, estimated through a linear combination of the input features (i.e., type 0, type 1, type 2, and duration) passed through a sigmoidal function, and \begin{math} \alpha, \beta_{1}, \beta_{2}, \beta_{3}, and \beta_{4} \end{math} are the parameters to be learned by model $M_{1}$.

\mbox{ }
\newline
Furthermore, a productive meeting might also depend on the psychological safety experienced by participants (i.e., whether they felt listened to). As we previously illustrated (\S\ref{sec:section3}), psychological factors such as inclusiveness~\cite{hbr_better_meetings, hbr_quality_experience}, dominance~\cite{romano2001meeting}, and comfort in sharing and contributing~\cite{hbr_tense, hbr_psychological_safety} are linked with people's overall experience of a meeting. To understand whether these factors also impact meetings' productivity, we formulated our second hypothesis as:\newline\newline\noindent\textbf{H\textsubscript{2}}:  Whether a meeting is productive depends on the meeting's \emph{type}, \emph{duration}, and on the \emph{psychological safety} experienced by participants. 

To investigate $H_{2}$, we used the z-score transformed ($\frac{x - \mu_{x}}{\sigma_{x}}$) of $Q_{psychological}$ reported by our participants (Figure~\ref{fig:freq_dist}b) along with meetings' \emph{duration}, and the three dummy variables, which reflect the meeting \emph{type}.

We built the $M_{2}$ logistic regression model (\ref{eq:M2}) to investigate $H_{2}$ by adding \emph{psychological safety} to the $M_{1}$ model. We used the binarized self-reported $Q_{productive}$ as the dependent variable:

\begin{gather}
  \shortintertext{\quad$ \Prb(\text{Q\textsubscript{productive}} = 1 \mid \text{type 0, type 1, type 2, duration, Q\textsubscript{psychological}}) = $} \frac{1}{1 + exp^-\textsuperscript{($\alpha$ \text{ + } $\beta_{1}$ $type 0$ \text{ + } $\beta_{2}$ $type 1$ \text{ + } $\beta_{3}$ $type2$ \text{ + } 
$\beta_{4}$ $duration$ \text{ + } 
$\beta_{5}$ $Q_{psychological}$ )}}, \label{eq:M2}
\end{gather}
where Pr is the posterior probability of an instance belonging to the positive class of $Q_{productive}$, estimated through a linear combination of the input features (i.e., type 0, type 1, type 2, duration, and $Q_{psychological}$) passed through a sigmoidal function, and \begin{math} \alpha, \beta_{1}, \beta_{2}, \beta_{3}, \beta_{4}, and \beta_{5} \end{math} are the parameters to be learned by model $M_{2}$.

\mbox{ }
\newline
Finally, the role of indoor environmental conditions into workplace's productivity has been the subject of several reports. For example, the lighting conditions were among the most popular perks\footnote{https://futureworkplace.com/} employees crave for at the workplace~\cite{hbr_office_light}. Additionally, studies have shown that stuffy and stale air offices reduce  productivity~\cite{allen2016associations, hbr_air_pollution, hbr_office_air}. Other studies  have reported that the actual environmental conditions might not always match the subjective perception of the physical space though~\cite{de2014measured,kang2017impact}. Therefore, we formulated our third hypothesis as:\newline\newline\noindent\textbf{H\textsubscript{3}}: Whether a meeting is productive depends on the meeting's
\emph{environmental conditions}, both actual and perceived conditions.

To investigate $H_{3}$, we used the z-score transformed self-reported $Q_{psychological}$ along with two metrics that capture the \emph{sensed pleasantness} and the \emph{not-sensed} one. The sensed pleasantness is a metric that captures the environmental conditions our ComFeel framework is able to capture. To built this metric, we resorted to the sensor readings obtained using our Gecko devices, and we did so in two steps. First, we used the indoor \emph{temperature} level (C\degree), \emph{gas resistance} (ohm), and \emph{light luminosity} (fc: foot-candle) (Figure~\ref{fig:freq_dist}f-h). For the sake of brevity, we refer to them as \emph{temp}, \emph{gas}, and \emph{light} respectively for the rest of our analysis. As the device was set to sample sensor readings every 60 seconds (\S\ref{subsec:gecko}), we aggregated the minutely obtained values across each meeting's duration, took their median value, and z-score ($\frac{x - \mu_{x}}{\sigma_{x}}$) transformed them to then make use of them in a logistic regression. We chose the median statistic as a more representative value due to its robustness against outliers, while the choice of using z-score transformations as opposed to raw values allowed us to incorporate the variance of the sensor-derived data. We then computed a logistic regression as:

\begin{figure*}
\captionsetup{labelfont=normalfont}
\centering
\subfloat[]{\includegraphics[width=0.2\textwidth]{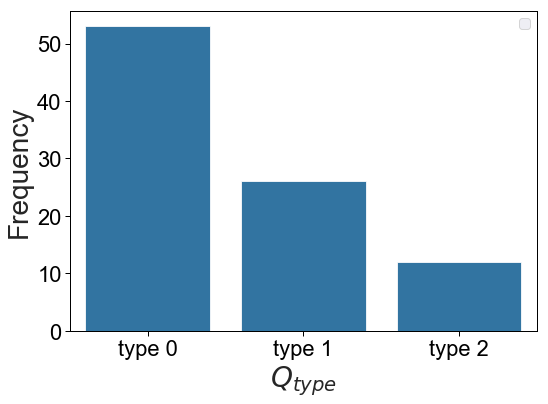}}\hfil
\subfloat[]{\includegraphics[width=0.2\textwidth]{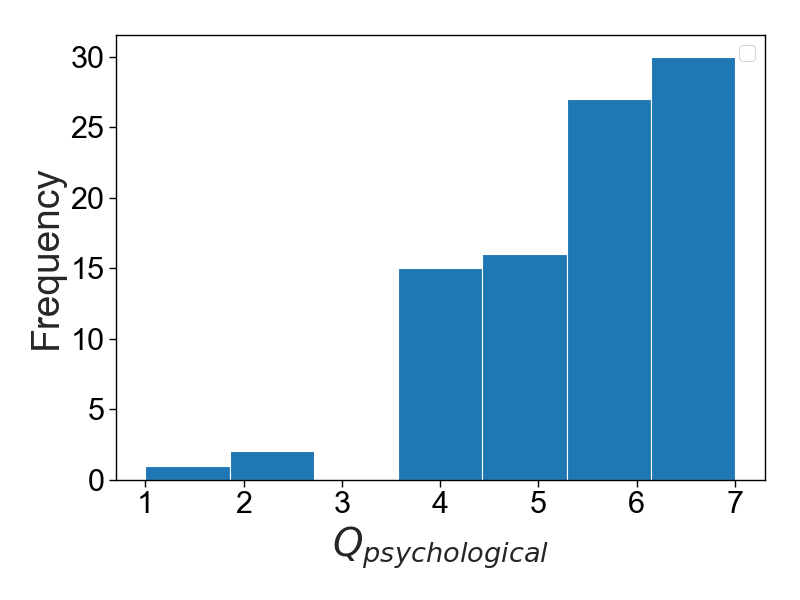}}\hfil
\subfloat[]{\includegraphics[width=0.2\textwidth]{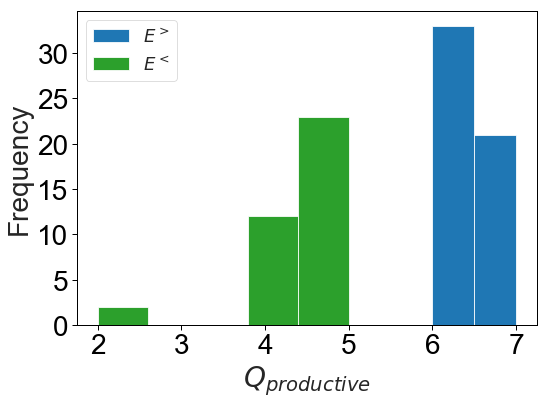}}\hfil
\subfloat[]{\includegraphics[width=0.2\textwidth]{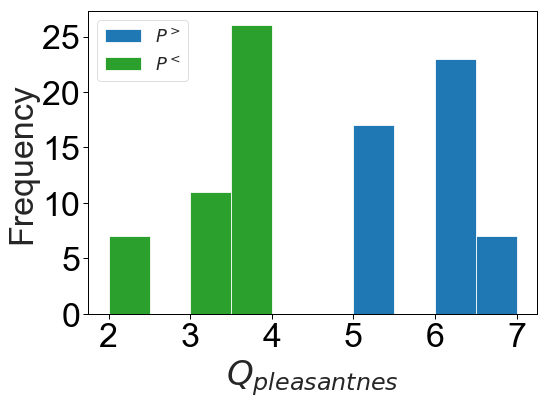}}\hfil

\subfloat[]{\includegraphics[width=0.2\textwidth]{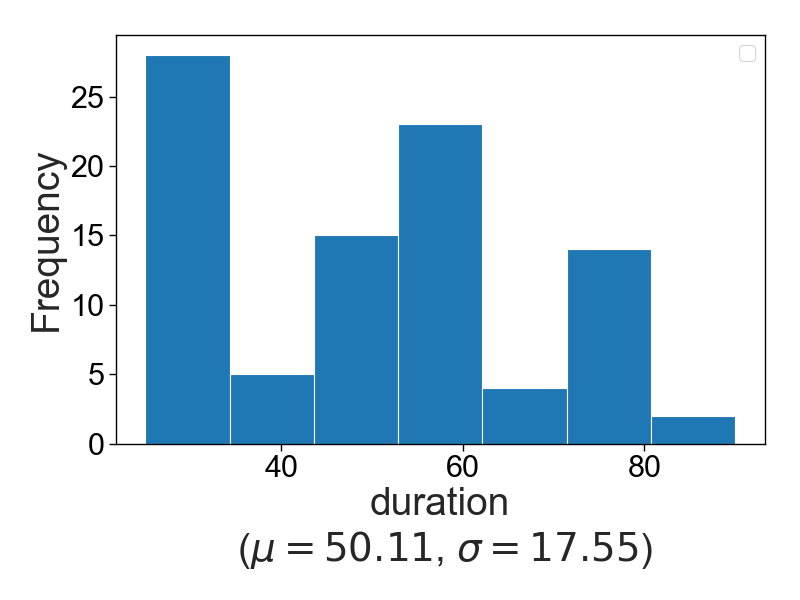}}\hfil
\subfloat[]{\includegraphics[width=0.2\textwidth]{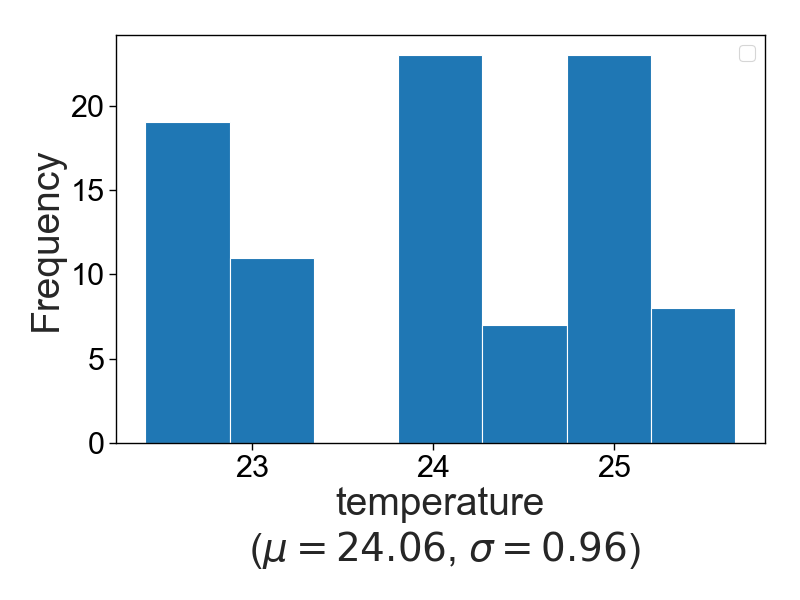}}\hfil
\subfloat[]{\includegraphics[width=0.2\textwidth]{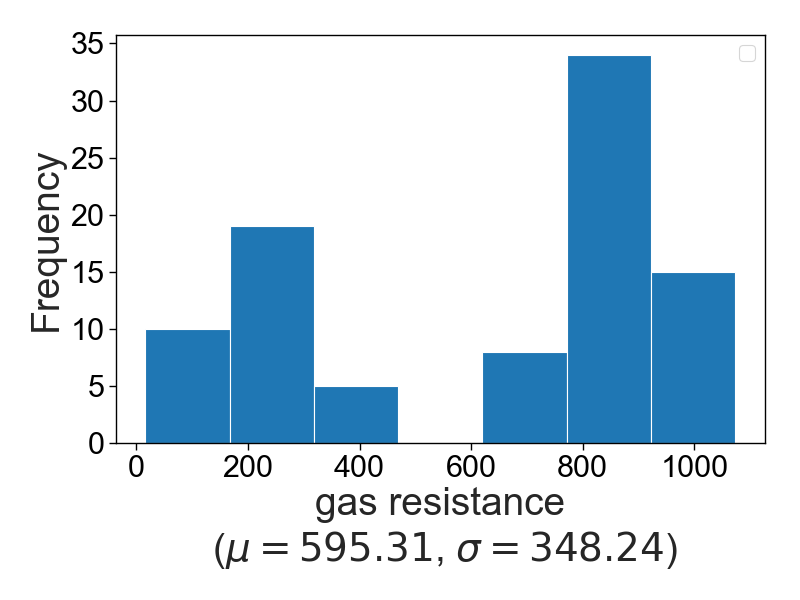}} \hfil
\subfloat[]{\includegraphics[width=0.2\textwidth]{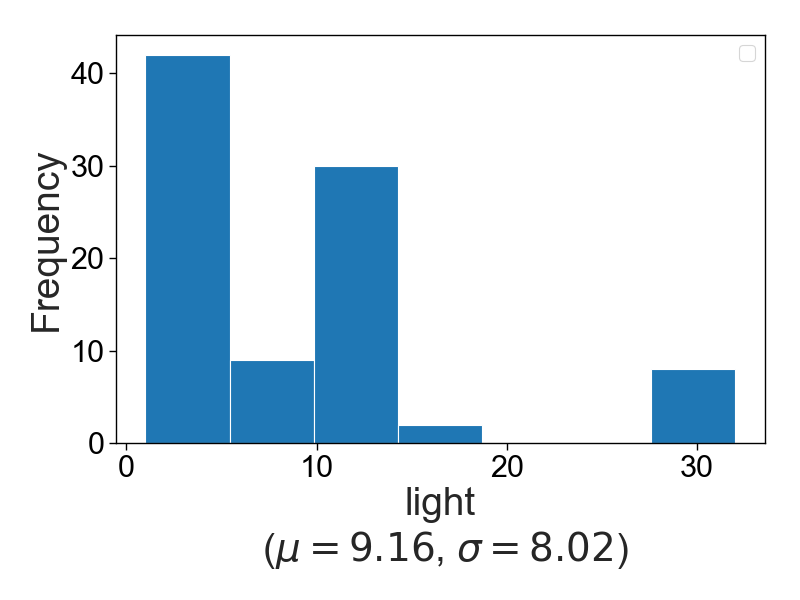}}\hfil 
\caption{Frequency distribution of variables used in the analysis. (a) $Q_{type}$; (b) $Q_{psychological}$; (c) $Q_{productive}$;  (d) $Q_{pleasantness}$; (e) duration; (f) temperature; (g) gas resistance; (h) light.} \label{fig:freq_dist}
\end{figure*}

\begin{equation}
\label{eq:1}
\operatorname
{Pr}(\text{$Q_{pleasantness}$}=1\mid\text{temp, gas, light}) = \frac{1}{1 + exp^-\textsuperscript{($\alpha$ \text{ + } $\beta_{1}$ $temp$ \text{ + } $\beta_{2}$ $gas$ \text{ + } $\beta_{3}$ $light$)}},
\end{equation}
where Pr is the posterior probability of being in the positive class of $Q_{pleasantness}$, and \begin{math} \alpha, \beta_{1}, \beta_{2}, and \beta_{3} \end{math} are the parameters to be learned. To binarize $Q_{pleasantness}$ (i.e., positive and negative class), we defined a set $P$ as the set of all participants' responses to $Q_{pleasantness}$, and we partitioned it into two groups such that $P = P^{>} \cup P^{<}$, where $P^{>}$ is the set of participants who reported higher levels of room pleasantness (above the median) defined as $P^{>} = \{{p \mid > \bar{p}\}}$, where $p$ is a participant's $Q_{pleasantness}$ response, and $P^{<}$ is the set of participants who reported lower levels of room pleasantness (below the median) defined as $P^{<} = \{{p \mid > \bar{p}\}}$, where $p$ is a participant's $Q_{pleasantness}$ response. Then, the $P^{>}$ responses were assigned to the positive class, while the $P^{<}$ ones to the negative (Figure~\ref{fig:freq_dist}d).

Secondly, using the learned parameters (i.e., $\alpha, \beta_{1}, \beta_{2}, and \beta_{3}$) from (\ref{eq:1}), we linearly combined them and computed \emph{sensed} as:
\begin{equation}
    \label{eq:2}
    sensed = (\alpha + \beta_{1}temp + \beta_{2}gas + \beta_{3}light),
\end{equation}
where $temp$, $gas$, and $light$ are the sensor readings obtained using our Gecko devices. We then computed \emph{z-sensed} as the z-score ($\frac{x - \mu_{x}}{\sigma_{x}}$) of \emph{sensed}.

The \emph{not-sensed} pleasantness metric captures the difference between the sensed and the self-reported pleasantness, and we computed it as:
\begin{equation}
    \text{z-not-sensed} = (\text{\emph{z-sensed}} - \text{\emph{z-$Q_{pleasantness}$})}
    \label{eq:gap}
\end{equation}
 The higher its value, the more our sensors overestimated what people felt, while the lower its value, the more our sensors underestimated what people felt.
 
We built the $M_{3}$ logistic regression model (\ref{eq:M3}) to address $H_{3}$ using the \emph{$Q_{psychological}$}, \emph{z-sensed}, and \emph{z-not-sensed}. 

\begin{gather}
  \shortintertext{\quad$ \Prb(\text{Q\textsubscript{productive}} = 1 \mid \text{Q\textsubscript{psychological}, z-sensed, z-not-sensed}) = $} \frac{1}{1 + exp^-\textsuperscript{($\alpha$ \text{ + } $\beta_{1}$ $Q_{psychological}$ \text{ + } $\beta_{2}$ z-sensed \text{ + } $\beta_{3}$ z-not-sensed  )}}, \label{eq:M3}
\end{gather}
where Pr is the posterior probability of being in the positive class of $Q_{productive}$, estimated through a linear combination of the input features (i.e., $Q_{psychological}$, z-sensed, and z-not-sensed) passed through a sigmoidal function, and \begin{math} \alpha, \beta_{1}, \beta_{2}, and \beta_{3} \end{math} are the parameters to be learned by model $M_{3}$.

\begin{figure*}
\captionsetup{labelfont=normalfont}
    \centering
    \subfloat[]{
        \includegraphics[width=0.7\textwidth ]{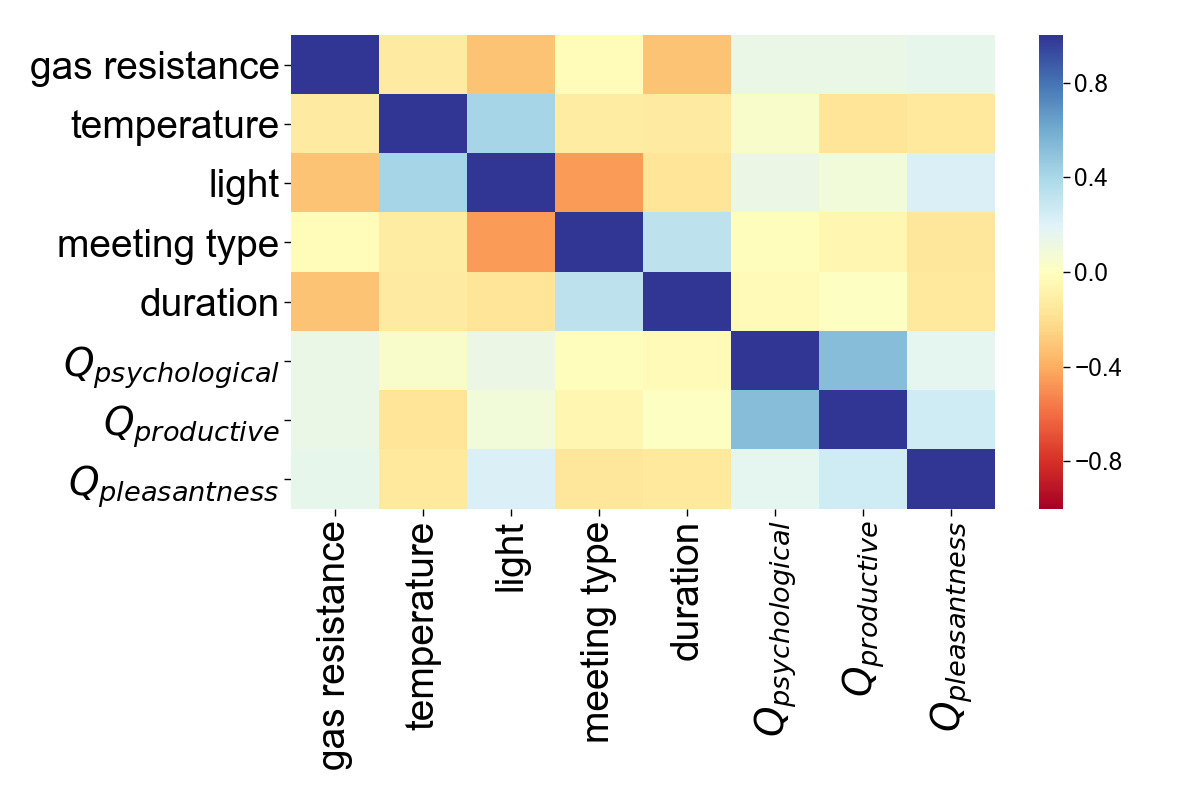}
        \label{fig:ada}
    }
    \caption{Cross-correlation matrix of the variables in our analyses. Overall, the independent variables are fairly orthogonal, while some pairs of variables such as light and gas resistance show weak to moderate correlation.}
    \label{fig:cross_corr}
\end{figure*}
\section{Results}
\label{sec:results}

To evaluate the performance of our models, we used the standard classification performance metrics. Specifically, we calculated the \emph{Area Under the Receiver Operating Characteristics (ROC) curve (AUC)} for our predictions. The ROC is a function of the true positive rate (TPR) against the false positive rate (FPR) of a classifier. Hence AUC will be larger for models that are good at predicting the positive class while not, at the same time, misclassifying the negative class. 

All three models' AUC scores and their significant predictors are shown in Table 5. We observed that $M_{1}$ predicts meetings' productivity slightly better than tossing a coin (AUC = 56\%), which led us to \emph{reject} \textbf{H\textsubscript{1}} based on the finding that whether a meeting was productive or not did not depend on the meeting's type or duration only. When adding the \emph{$Q_{psychological}$} into $M_{2}$ model, however, the performance drastically improved, reaching an AUC of 82\%. This suggests that participants' perception on whether a meeting was productive is mediated by the psychological safety that they experienced, in line with previous findings on the importance of psychological factors that allows for effective coordination and  cooperation in a meeting, leading us to \emph{accept} \textbf{H\textsubscript{2}}. Finally, we found that the performance of our $M_{3}$ model further improves reaching up to 86\%. This translates into saying that the environmental conditions, both actual and perceived ones, further mediate people's perception on whether a meeting was productive. Therefore, we \emph{accept} \textbf{H\textsubscript{3}}, and we attempt to explain these observations by interpreting $M_{3}$'s model coefficients.

\begin{table}
\label{tbl:results}
\captionsetup{labelfont=normalfont}
    \caption{Logistic regression coefficients. $M_{0}$ model's coefficients were used to predict \emph{sensed} pleasantness; $M_{1}, M_{2} \text{ and } M_{3}$ tested hypotheses $H_{1}, H_{2} \text{, and } H_{3}$ respectively. Statistically significant predictors at $< .05$ are marked in bold. (-) indicates that a model was fit without that predictor.}
    \begin{minipage}[t]{.5\linewidth}
      \centering
            \begin{tabular}{lc}
    \toprule
    & \textbf{M\textsubscript{0}} 
     \\
    \hline
    \textbf{Predictors} &  
      \\
    \midrule
    Intercept ($\alpha$) & -0.576  \\
    temp & \textbf{-0.630}\\
    gas & \textbf{1.121} \\
    light & \textbf{0.752}\\
     \midrule
    \textbf{AUC score} & 77.08\% \\
  \bottomrule
\end{tabular}
    \end{minipage}%
    \begin{minipage}[t]{.5\linewidth}
      \centering
        \begin{tabular}{lccc}
    \toprule
    & \textbf{M\textsubscript{1}} 
    & \textbf{M\textsubscript{2}} 
    & \textbf{M\textsubscript{3}} \\
    \hline
    \textbf{Predictors}  &  & &
      \\
    \midrule
   Intercept ($\alpha$)  & -0.089  & -0.613 & -0.646 \\
    type 0 & -0.165 & 0.051 & -\\
    type 1 & -0.089 & -0.635 & -\\
    type 2&  0.064 & -0.028 & -\\
    duration & -0.009 & -0.001 & -\\
    $Q_{psychological}$ & - & \textbf{1.526}  & \textbf{1.389}\\
    z-sensed &-  & -  & \textbf{1.005} \\
    z-not-sensed & - &  -&  -0.256\\
     \midrule
    \textbf{AUC score} & 55.72\% & 82.56\% & 85.84\%\\
  \bottomrule
\end{tabular}
    \end{minipage} 
\end{table}

Additionally, we run a fourth model by including all the seven predictors and observed an AUC of 86.6\%. This yielded very little gain in AUC compared to $M_{3}$. Further, to explore whether there are any interactions between the variables in our analyses, we measured whether pairs of the independent variables correlated with each other. To do so, we computed the Spearman's rank cross-correlation matrix among all pairs (Figure~\ref{fig:cross_corr}). Some pairs such as light and gas resistance, and light and meeting type (as certain types of meeting, such as status updates, tended to happen more often in the morning) show weak to moderate correlations, but overall our independent variables are fairly orthogonal.

To get an upper bound of the predictive difference corresponding to a unit difference in each predictor, we use the `divide by 4' rule~\cite{gelman2007analytical} (p. 82). This says that, in a logistic regression, by dividing each predictor's coefficient by 4 one can draw conclusions for 1 unit increase in each predictor. Simply put, we can quantify the percentage of increase or decrease in the probability of the dependent variable to be in the positive class.

\subsection{The Known Unknowns of Productivity}
By applying this rule, we inspected $M_{3}$'s model predictors' coefficients and found that the probability of a meeting perceived as a productive one is increased by 35\% every one standard deviation increase in the \emph{$Q_{psychological}$}. This corroborates previous findings, which highlighted the importance of inclusiveness~\cite{hbr_better_meetings, hbr_quality_experience}, dominance~\cite{romano2001meeting}, and comfort in sharing and contributing~\cite{hbr_tense, hbr_psychological_safety} in meetings. Furthermore, we found that the probability of a meeting perceived as a productive one is increased by 25\% every one standard deviation increase in the \emph{z-sensed} pleasantness. This suggests that environmental conditions also impact meetings' productivity, supporting this work's overall hypothesis that environmental conditions impact not only long-term productivity (as suggested by previous work~\cite{hbr_office_light, allen2016associations, hbr_air_pollution, hbr_office_air}) but also short-term and immediate one. We also found that the probability of perceiving meetings to be productive is increased by 6\% every one standard deviation decrease in the part of people's self-reported \emph{$Q_{pleasantness}$} that our sensors could not capture (encoded into \emph{z-not-sensed}), confirming that ComFeel captured most of the variability in the environmental conditions  people experienced (adding external validity to our framework's evaluation) and, at the same time, suggesting the orthogonal importance of asking a simple question even in the presence of an effective sensing infrastructure.
\section{Discussion}
\label{sec:discussion}

Our operationalization of the concept of sensorial pleasantness and productivity allowed us to study, for the first time, the short term impact of environmental conditions on meetings' productivity. By analyzing an online survey, which we administered to 363 participants, we unveiled three orthogonal dimensions that capture meetings' psychological experience:  (a) \emph{psychological comfort}, (b) \emph{productivity}, and (c) \emph{room pleasantness}.

Using the collected data obtained from our ComFeel framework and employee's self-reported experiences during their meetings, we investigated the interplay of environmental conditions with productivity. We first captured the sensed pleasantness by employing a machine learning method on three environmental conditions (i.e., temperature, light, and gas resistance), moving beyond thermal comfort previous work focused on~\cite{song2019human}. We built a model that learned the linear combination of these factors, and we found that gas resistance was the most predictive one, followed by light and temperature. We then built three models to test our three hypotheses by incrementally introducing variables. In our final $M_{3}$ model, we found that, on average, the probability of a meeting perceived as a productive one increased by 35\% for each standard deviation increase in the psychological safety participants experience. This result confirms previous findings reported in literature~\cite{hbr_better_meetings, hbr_quality_experience, romano2001meeting, hbr_tense, hbr_psychological_safety}, and suggests that meetings' productivity is mediated by whether participants felt listened to and were able to freely contribute. Additionally, we found that the very same probability increased by as much as 25\% for each standard deviation increase in room pleasantness, suggesting the short-term impact of environmental conditions in meetings' productivity.

\subsection{Implications}
The theoretical implications of our work concern the fields of Management and Organizational Science. By analyzing our online questionnaire, we found that productivity, psychological safety, and room pleasantness capture, to a great extent, people's experience of meetings. This finding corroborates previous ones, which highlighted the importance of inclusiveness~\cite{hbr_better_meetings, hbr_quality_experience}, dominance~\cite{romano2001meeting}, and comfort in sharing and contributing in meetings~\cite{hbr_tense, hbr_psychological_safety}. All these aspects  could be potentially captured by simply  administering only three questions, 
making it possible for a large organization to understand their employees' overall experience during meetings.

Practically, our work speaks to the broader Ubicomp community. We found that particular environmental properties were more significant than others to predict perceived environmental conditions, and this could drive the focus of the research on sensing. Specifically, among the three environmental properties used to capture room pleasantness, we found that  gas resistance was the most significant one, while lighting conditions and temperature had a second-order, yet significant and considerable effect. A finding also aligned with previous work that showed that high exposures to volatile compounds (which gas resistance captures) were associated with a decline in cognitive functions~\cite{allen2016associations}. As of immediate practical use, our ComFeel framework is an easy-to-deploy infrastructure that could be used across any meeting room within an organization. Its flexible design, data reliability (off-the-shelf environmental sensors), and low manufacturing costs make it an appealing device for future deployments across any organization, independently to its size. More broadly, our ComFeel framework could further facilitate research on Human-Building Interaction (HBI); an emerging area aiming at unifying HCI research in the built environment~\cite{alavi2019introduction}. Building on HBI's mission, we foresee to be able to answer questions related to human comfort (consider situations in shared offices where comfort needs to be agreed upon, even with basic interventions such as opening a window to more automated ones like adjusting automatically the thermostat setting) and the broader social dimensions of the workplace~\cite{alavi2017comfort, alavi2016future}. Beyond the workplace, ComFeel could also be utilized in smart factories, which are considered an integral part of the fourth industrial revolution~\cite{wang2016implementing}. The very same technology used here might well be used in an industrial setting to allow companies to optimize for their workers' productivity and well-being in real time.

\subsection{Limitations and Future Work}
The current work is subject to three main limitations that call for further work. The first limitation concerns the online survey. While we adopted a rigorous statistical approach to derive the three main principal components, we eliminated seventeen questions from the initial set that account for 37.69\% of the variance. As previously stated though (\S\ref{sec:section3}), a set of three questions that accounts for 60 percent of the total variance can be deemed as satisfactory~\cite{hair1998multivariate}. Additionally, the choice of the main three principal components was also reinforced by the practicalities of asking three questions only, making it attractive for deployments in a real application. Although a questionnaire's length and its response rate is often debatable~\cite{rolstad2011response}, a shorter set of questions makes our application practical to use. The second limitation is about the sample size used in the analyses. Despite the fact that we obtained a fairly large amount of raw environmental readings, and hundreds of self-reported answers, a larger scale deployment would further benefit our analyses beyond data obtained at the facilities of one building. Immediate future work includes the deployment of our ComFeel framework in a broader set of meetings within our office's environment and then across the entire organization, thus enabling us to investigate other factors such as people's role within the company (e.g., individual contributors, managers, executives). The third limitation concerns the data collection and, particularly, the off-the-shelf environmental sensors used to obtain the readings. Despite these sensors are commercial ones, and were validated separately, future directions include the integration of additional and alternative vendors into our framework. Additionally, in future, we plan to extend our devices by incorporating additional sensors that capture a wider set of environmental factors such as noise or humidity. Along these lines, our models make use of the perceived productivity levels that our participants reported, whereas alternative ways of obtaining objective measurements of meetings' productivity (e.g., number of completed agenda items) could also be investigated in future work.
\section{Conclusion}
\label{sec:conclusion}

We studied the short term impact of indoor environmental quality on productivity at one of the most common daily activity at work: \emph{meetings}. We operationalized the concept of sensorial pleasantness and productivity in meetings by analyzing an online questionnaire, and identified three factors that capture people's experience of meetings, namely psychological comfort, productivity, and room pleasantness. Building on these insights, we measured room pleasantness using ComFeel, an indoor environmental sensing infrastructure we developed that captures light, temperature, and gas resistance readings through our Gecko devices. We deployed our ComFeel across 29 real-world meetings, which spanned over a period of one month, and collected 1373 minutes of environmental sensor readings. For each of these meetings, we also collected whether each participant felt the meeting to have been productive, the setting to be  psychologically safe, and the meeting room to be pleasant. As one expects, we found that, on average, the probability of a meeting being productive increased by 35\% for each standard deviation increase in the psychological safety participants experienced. Importantly, the very same probability increased by as much as 25\% for each standard deviation increase in room pleasantness. These results suggest that environmental conditions do matter not only in the long term but also in the short term so much so that significant differences in productivity were observed even within the constrained space of individual meetings.

\section*{Acknowledgments}
We thank Philip Derrick for his contributions in the development of the Gecko devices. Thanks also to our participants and anonymous reviewers for their comments and feedback on this manuscript.

\bibliographystyle{ACM-Reference-Format}
\bibliography{main}


\begin{thebibliography}{79}


\ifx \showCODEN    \undefined \def \showCODEN     #1{\unskip}     \fi
\ifx \showDOI      \undefined \def \showDOI       #1{#1}\fi
\ifx \showISBNx    \undefined \def \showISBNx     #1{\unskip}     \fi
\ifx \showISBNxiii \undefined \def \showISBNxiii  #1{\unskip}     \fi
\ifx \showISSN     \undefined \def \showISSN      #1{\unskip}     \fi
\ifx \showLCCN     \undefined \def \showLCCN      #1{\unskip}     \fi
\ifx \shownote     \undefined \def \shownote      #1{#1}          \fi
\ifx \showarticletitle \undefined \def \showarticletitle #1{#1}   \fi
\ifx \showURL      \undefined \def \showURL       {\relax}        \fi
\providecommand\bibfield[2]{#2}
\providecommand\bibinfo[2]{#2}
\providecommand\natexlab[1]{#1}
\providecommand\showeprint[2][]{arXiv:#2}

\bibitem[\protect\citeauthoryear{Alavi, Churchill, Wiberg, Lalanne, Dalsgaard,
  Fatah~gen Schieck, and Rogers}{Alavi et~al\mbox{.}}{2019}]%
        {alavi2019introduction}
\bibfield{author}{\bibinfo{person}{Hamed~S Alavi}, \bibinfo{person}{Elizabeth~F
  Churchill}, \bibinfo{person}{Mikael Wiberg}, \bibinfo{person}{Denis Lalanne},
  \bibinfo{person}{Peter Dalsgaard}, \bibinfo{person}{Ava Fatah~gen Schieck},
  {and} \bibinfo{person}{Yvonne Rogers}.} \bibinfo{year}{2019}\natexlab{}.
\newblock \showarticletitle{Introduction to Human-Building Interaction (HBI)
  Interfacing HCI with Architecture and Urban Design}.
\newblock \bibinfo{journal}{\emph{ACM Transactions on Computer-Human
  Interaction (TOCHI)}} \bibinfo{volume}{26}, \bibinfo{number}{2}
  (\bibinfo{year}{2019}).
\newblock


\bibitem[\protect\citeauthoryear{Alavi, Lalanne, Nembrini, Churchill, Kirk, and
  Moncur}{Alavi et~al\mbox{.}}{2016}]%
        {alavi2016future}
\bibfield{author}{\bibinfo{person}{Hamed~S Alavi}, \bibinfo{person}{Denis
  Lalanne}, \bibinfo{person}{Julien Nembrini}, \bibinfo{person}{Elizabeth
  Churchill}, \bibinfo{person}{David Kirk}, {and} \bibinfo{person}{Wendy
  Moncur}.} \bibinfo{year}{2016}\natexlab{}.
\newblock \showarticletitle{Future of Human-Building Interaction}. In
  \bibinfo{booktitle}{\emph{Proceedings of the 2016 CHI Conference Extended
  Abstracts on Human Factors in Computing Systems}}.
  \bibinfo{pages}{3408--3414}.
\newblock


\bibitem[\protect\citeauthoryear{Alavi, Verma, Papinutto, and Lalanne}{Alavi
  et~al\mbox{.}}{2017}]%
        {alavi2017comfort}
\bibfield{author}{\bibinfo{person}{Hamed~S Alavi}, \bibinfo{person}{Himanshu
  Verma}, \bibinfo{person}{Michael Papinutto}, {and} \bibinfo{person}{Denis
  Lalanne}.} \bibinfo{year}{2017}\natexlab{}.
\newblock \showarticletitle{Comfort: A Coordinate of User Experience in
  Interactive Built Environments}. In \bibinfo{booktitle}{\emph{IFIP Conference
  on Human-Computer Interaction}}. Springer, \bibinfo{pages}{247--257}.
\newblock


\bibitem[\protect\citeauthoryear{Ali, Cot{\'e}, Heidarinejad, and Stephens}{Ali
  et~al\mbox{.}}{2019}]%
        {ali2019elemental}
\bibfield{author}{\bibinfo{person}{Akram~Syed Ali},
  \bibinfo{person}{Christopher Cot{\'e}}, \bibinfo{person}{Mohammad
  Heidarinejad}, {and} \bibinfo{person}{Brent Stephens}.}
  \bibinfo{year}{2019}\natexlab{}.
\newblock \showarticletitle{Elemental: An Open-Source Wireless Hardware and
  Software Platform for Building Energy and Indoor Environmental Monitoring and
  Control}.
\newblock \bibinfo{journal}{\emph{Sensors}} \bibinfo{volume}{19},
  \bibinfo{number}{18} (\bibinfo{year}{2019}), \bibinfo{pages}{4017}.
\newblock


\bibitem[\protect\citeauthoryear{Allen}{Allen}{2017}]%
        {hbr_office_air}
\bibfield{author}{\bibinfo{person}{Joseph~G. Allen}.}
  \bibinfo{year}{2017}\natexlab{}.
\newblock \showarticletitle{Stale Office Air Is Making You Less Productive}.
\newblock \bibinfo{journal}{\emph{Harvard Business Review}}
  (\bibinfo{year}{2017}).
\newblock


\bibitem[\protect\citeauthoryear{Allen, MacNaughton, Satish, Santanam,
  Vallarino, and Spengler}{Allen et~al\mbox{.}}{2016}]%
        {allen2016associations}
\bibfield{author}{\bibinfo{person}{Joseph~G Allen}, \bibinfo{person}{Piers
  MacNaughton}, \bibinfo{person}{Usha Satish}, \bibinfo{person}{Suresh
  Santanam}, \bibinfo{person}{Jose Vallarino}, {and} \bibinfo{person}{John~D
  Spengler}.} \bibinfo{year}{2016}\natexlab{}.
\newblock \showarticletitle{Associations of Cognitive Function Scores with
  Carbon Dioxide, Ventilation, and Volatile Organic Compound Exposures in
  Office Workers: A Controlled Exposure Study of Green and Conventional Office
  Environments}.
\newblock \bibinfo{journal}{\emph{Environmental Health Perspectives}}
  \bibinfo{volume}{124}, \bibinfo{number}{6} (\bibinfo{year}{2016}),
  \bibinfo{pages}{805--812}.
\newblock


\bibitem[\protect\citeauthoryear{Aseniero, Constantinides, Joglekar, Zhou, and
  Quercia}{Aseniero et~al\mbox{.}}{2020}]%
        {meetcues}
\bibfield{author}{\bibinfo{person}{Bon~Adriel Aseniero},
  \bibinfo{person}{Marios Constantinides}, \bibinfo{person}{Sagar Joglekar},
  \bibinfo{person}{Ke Zhou}, {and} \bibinfo{person}{Daniele Quercia}.}
  \bibinfo{year}{2020}\natexlab{}.
\newblock \showarticletitle{MeetCues: Supporting Online Meetings Experience}.
  In \bibinfo{booktitle}{\emph{Proc. of the IEEE Visualization Conference
  (VIS)}}. IEEE.
\newblock


\bibitem[\protect\citeauthoryear{Axtell}{Axtell}{2016}]%
        {hbr_better_meetings}
\bibfield{author}{\bibinfo{person}{Paul Axtell}.}
  \bibinfo{year}{2016}\natexlab{}.
\newblock \showarticletitle{6 Reasons to Get Better at Leading Meetings}.
\newblock \bibinfo{journal}{\emph{Harvard Business Review}}
  (\bibinfo{year}{2016}).
\newblock


\bibitem[\protect\citeauthoryear{Axtell}{Axtell}{2017}]%
        {hbr_quality_experience}
\bibfield{author}{\bibinfo{person}{Paul Axtell}.}
  \bibinfo{year}{2017}\natexlab{}.
\newblock \showarticletitle{How to Design Meetings Your Team Will Want to
  Attend}.
\newblock \bibinfo{journal}{\emph{Harvard Business Review}}
  (\bibinfo{year}{2017}).
\newblock


\bibitem[\protect\citeauthoryear{Axtell}{Axtell}{2018}]%
        {hbr_psychological_safety}
\bibfield{author}{\bibinfo{person}{Paul Axtell}.}
  \bibinfo{year}{2018}\natexlab{}.
\newblock \showarticletitle{How to Respond When You're Put on the Spot in a
  Meeting}.
\newblock \bibinfo{journal}{\emph{Harvard Business Review}}
  (\bibinfo{year}{2018}).
\newblock


\bibitem[\protect\citeauthoryear{Bader, Voit, Le, Wo{\'z}niak, Henze, and
  Schmidt}{Bader et~al\mbox{.}}{2019}]%
        {bader2019windowwall}
\bibfield{author}{\bibinfo{person}{Patrick Bader}, \bibinfo{person}{Alexandra
  Voit}, \bibinfo{person}{Huy~Viet Le}, \bibinfo{person}{Pawe{\l}~W
  Wo{\'z}niak}, \bibinfo{person}{Niels Henze}, {and} \bibinfo{person}{Albrecht
  Schmidt}.} \bibinfo{year}{2019}\natexlab{}.
\newblock \showarticletitle{WindowWall: Towards Adaptive Buildings with
  Interactive Windows as Ubiquitous Displays}.
\newblock \bibinfo{journal}{\emph{ACM Transactions on Computer-Human
  Interaction (TOCHI)}} \bibinfo{volume}{26}, \bibinfo{number}{2}
  (\bibinfo{year}{2019}), \bibinfo{pages}{1--42}.
\newblock


\bibitem[\protect\citeauthoryear{Bako-Biro, Wargocki, Weschler, and
  Fanger}{Bako-Biro et~al\mbox{.}}{2004}]%
        {bako2004effects}
\bibfield{author}{\bibinfo{person}{Zsolt Bako-Biro}, \bibinfo{person}{Pawel
  Wargocki}, \bibinfo{person}{Charles~J Weschler}, {and}
  \bibinfo{person}{Povl~Ole Fanger}.} \bibinfo{year}{2004}\natexlab{}.
\newblock \showarticletitle{Effects of pollution from personal computers on
  perceived air quality, SBS symptoms and productivity in offices}.
\newblock \bibinfo{journal}{\emph{Indoor air}} \bibinfo{volume}{14},
  \bibinfo{number}{3} (\bibinfo{year}{2004}), \bibinfo{pages}{178--187}.
\newblock


\bibitem[\protect\citeauthoryear{Barrick and Mount}{Barrick and Mount}{1991}]%
        {barrick1991big}
\bibfield{author}{\bibinfo{person}{Murray~R Barrick} {and}
  \bibinfo{person}{Michael~K Mount}.} \bibinfo{year}{1991}\natexlab{}.
\newblock \showarticletitle{The Big Five Personality Dimensions and Job
  Performance: A Meta-analysis}.
\newblock \bibinfo{journal}{\emph{Personnel psychology}} \bibinfo{volume}{44},
  \bibinfo{number}{1} (\bibinfo{year}{1991}), \bibinfo{pages}{1--26}.
\newblock


\bibitem[\protect\citeauthoryear{Burge}{Burge}{2004}]%
        {burge2004sick}
\bibfield{author}{\bibinfo{person}{P~Sherwood Burge}.}
  \bibinfo{year}{2004}\natexlab{}.
\newblock \showarticletitle{Sick building syndrome}.
\newblock \bibinfo{journal}{\emph{Occupational and Environmental Medicine}}
  \bibinfo{volume}{61}, \bibinfo{number}{2} (\bibinfo{year}{2004}),
  \bibinfo{pages}{185--190}.
\newblock


\bibitem[\protect\citeauthoryear{Cattell}{Cattell}{1966}]%
        {cattell1966scree}
\bibfield{author}{\bibinfo{person}{Raymond~B Cattell}.}
  \bibinfo{year}{1966}\natexlab{}.
\newblock \showarticletitle{The Scree Test for the Number of Factors}.
\newblock \bibinfo{journal}{\emph{Multivariate Behavioral Research}}
  \bibinfo{volume}{1}, \bibinfo{number}{2} (\bibinfo{year}{1966}),
  \bibinfo{pages}{245--276}.
\newblock


\bibitem[\protect\citeauthoryear{Clements-Croome and Baizhan}{Clements-Croome
  and Baizhan}{2000}]%
        {clements2000productivity}
\bibfield{author}{\bibinfo{person}{Derek Clements-Croome} {and}
  \bibinfo{person}{Li Baizhan}.} \bibinfo{year}{2000}\natexlab{}.
\newblock \showarticletitle{Productivity and indoor environment}. In
  \bibinfo{booktitle}{\emph{Proceedings of Healthy Buildings}},
  Vol.~\bibinfo{volume}{1}. \bibinfo{pages}{629--634}.
\newblock


\bibitem[\protect\citeauthoryear{Cohen}{Cohen}{2016}]%
        {hbr_cohen_start}
\bibfield{author}{\bibinfo{person}{Jordan Cohen}.}
  \bibinfo{year}{2016}\natexlab{}.
\newblock \showarticletitle{Use Subtle Cues to Encourage Better Meetings}.
\newblock \bibinfo{journal}{\emph{Harvard Business Review}}
  (\bibinfo{year}{2016}).
\newblock


\bibitem[\protect\citeauthoryear{De~Giuli, Zecchin, Corain, and
  Salmaso}{De~Giuli et~al\mbox{.}}{2014}]%
        {de2014measured}
\bibfield{author}{\bibinfo{person}{Valeria De~Giuli}, \bibinfo{person}{Roberto
  Zecchin}, \bibinfo{person}{Livio Corain}, {and} \bibinfo{person}{Luigi
  Salmaso}.} \bibinfo{year}{2014}\natexlab{}.
\newblock \showarticletitle{Measured and Perceived Environmental Comfort: Field
  Monitoring in an Italian School}.
\newblock \bibinfo{journal}{\emph{Applied Ergonomics}} \bibinfo{volume}{45},
  \bibinfo{number}{4} (\bibinfo{year}{2014}), \bibinfo{pages}{1035--1047}.
\newblock


\bibitem[\protect\citeauthoryear{De~Giuli, Zecchin, Salmaso, Corain, and
  De~Carli}{De~Giuli et~al\mbox{.}}{2013}]%
        {de2013measured}
\bibfield{author}{\bibinfo{person}{Valeria De~Giuli}, \bibinfo{person}{Roberto
  Zecchin}, \bibinfo{person}{Luigi Salmaso}, \bibinfo{person}{Livio Corain},
  {and} \bibinfo{person}{Michele De~Carli}.} \bibinfo{year}{2013}\natexlab{}.
\newblock \showarticletitle{Measured and Perceived Indoor Environmental
  Quality: Padua Hospital Case Study}.
\newblock \bibinfo{journal}{\emph{Building and Environment}}
  \bibinfo{volume}{59} (\bibinfo{year}{2013}), \bibinfo{pages}{211--226}.
\newblock


\bibitem[\protect\citeauthoryear{Fabrigar, Wegener, MacCallum, and
  Strahan}{Fabrigar et~al\mbox{.}}{1999}]%
        {fabrigar1999evaluating}
\bibfield{author}{\bibinfo{person}{Leandre~R Fabrigar},
  \bibinfo{person}{Duane~T Wegener}, \bibinfo{person}{Robert~C MacCallum},
  {and} \bibinfo{person}{Erin~J Strahan}.} \bibinfo{year}{1999}\natexlab{}.
\newblock \showarticletitle{Evaluating the Use of Exploratory Factor Analysis
  in Psychological Research}.
\newblock \bibinfo{journal}{\emph{Psychological Methods}} \bibinfo{volume}{4},
  \bibinfo{number}{3} (\bibinfo{year}{1999}), \bibinfo{pages}{272}.
\newblock


\bibitem[\protect\citeauthoryear{Fan, Miller, Park, Winward, Christensen,
  Grotevant, and Tai}{Fan et~al\mbox{.}}{2006}]%
        {fan2006exploratory}
\bibfield{author}{\bibinfo{person}{Xitao Fan}, \bibinfo{person}{Brent~C
  Miller}, \bibinfo{person}{Kyung-Eun Park}, \bibinfo{person}{Bryan~W Winward},
  \bibinfo{person}{Mathew Christensen}, \bibinfo{person}{Harold~D Grotevant},
  {and} \bibinfo{person}{Robert~H Tai}.} \bibinfo{year}{2006}\natexlab{}.
\newblock \showarticletitle{An Exploratory Study About Inaccuracy and
  Invalidity in Adolescent Self-report Surveys}.
\newblock \bibinfo{journal}{\emph{Field Methods}} \bibinfo{volume}{18},
  \bibinfo{number}{3} (\bibinfo{year}{2006}), \bibinfo{pages}{223--244}.
\newblock


\bibitem[\protect\citeauthoryear{Fanger}{Fanger}{1988}]%
        {fanger1988introduction}
\bibfield{author}{\bibinfo{person}{P~Ole Fanger}.}
  \bibinfo{year}{1988}\natexlab{}.
\newblock \showarticletitle{Introduction of the Olf and the Decipol Units to
  Quantify Air Pollution Perceived by Humans Indoors and Outdoors}.
\newblock \bibinfo{journal}{\emph{Energy and Buildings}} \bibinfo{volume}{12},
  \bibinfo{number}{1} (\bibinfo{year}{1988}), \bibinfo{pages}{1--6}.
\newblock


\bibitem[\protect\citeauthoryear{Fisk}{Fisk}{2000}]%
        {fisk2000health}
\bibfield{author}{\bibinfo{person}{William~J Fisk}.}
  \bibinfo{year}{2000}\natexlab{}.
\newblock \showarticletitle{Health and Productivity Gains from Better Indoor
  Environments And Their Relationship With Building Energy Efficiency}.
\newblock \bibinfo{journal}{\emph{Annual Review of Energy and the Environment}}
  \bibinfo{volume}{25}, \bibinfo{number}{1} (\bibinfo{year}{2000}),
  \bibinfo{pages}{537--566}.
\newblock


\bibitem[\protect\citeauthoryear{Gelman and Hill}{Gelman and Hill}{2007}]%
        {gelman2007analytical}
\bibfield{author}{\bibinfo{person}{A Gelman} {and} \bibinfo{person}{J Hill}.}
  \bibinfo{year}{2007}\natexlab{}.
\newblock \showarticletitle{Analytical Methods for Social Research}.
\newblock \bibinfo{journal}{\emph{Data analysis using regression and
  multilevel/hierarchical models}} (\bibinfo{year}{2007}).
\newblock


\bibitem[\protect\citeauthoryear{Goff-Dupont}{Goff-Dupont}{2018}]%
        {meeting_thinkgrowth}
\bibfield{author}{\bibinfo{person}{Sarah Goff-Dupont}.}
  \bibinfo{year}{2018}\natexlab{}.
\newblock \bibinfo{booktitle}{\emph{6 types of meetings that are actually
  worthwhile}}.
\newblock
\urldef\tempurl%
\url{https://thinkgrowth.org/6-types-of-meetings-that-are-actually-worthwhile-432877707493}
\showURL{%
\tempurl}


\bibitem[\protect\citeauthoryear{Grenny}{Grenny}{2017}]%
        {hbr_tense}
\bibfield{author}{\bibinfo{person}{Joseph Grenny}.}
  \bibinfo{year}{2017}\natexlab{}.
\newblock \showarticletitle{How to Save a Meeting That's Gotten Tense}.
\newblock \bibinfo{journal}{\emph{Harvard Business Review}}
  (\bibinfo{year}{2017}).
\newblock


\bibitem[\protect\citeauthoryear{Hair, Black, Babin, Anderson, Tatham,
  et~al\mbox{.}}{Hair et~al\mbox{.}}{1998}]%
        {hair1998multivariate}
\bibfield{author}{\bibinfo{person}{Joseph~F Hair}, \bibinfo{person}{William~C
  Black}, \bibinfo{person}{Barry~J Babin}, \bibinfo{person}{Rolph~E Anderson},
  \bibinfo{person}{Ronald~L Tatham}, {et~al\mbox{.}}}
  \bibinfo{year}{1998}\natexlab{}.
\newblock \bibinfo{booktitle}{\emph{Multivariate Data Analysis}}.
  Vol.~\bibinfo{volume}{5}.
\newblock \bibinfo{publisher}{Prentice Hall}.
\newblock


\bibitem[\protect\citeauthoryear{Hantani, Ikaga, Murakami, and Kameda}{Hantani
  et~al\mbox{.}}{2009}]%
        {hantani2009effect}
\bibfield{author}{\bibinfo{person}{Eriko Hantani}, \bibinfo{person}{Toshiharu
  Ikaga}, \bibinfo{person}{Shuzo Murakami}, {and} \bibinfo{person}{Ken~Ichi
  Kameda}.} \bibinfo{year}{2009}\natexlab{}.
\newblock \showarticletitle{Effect of Indoor Environmental Quality on
  Performance and Satisfaction in Call-center}. In
  \bibinfo{booktitle}{\emph{9th International Healthy Buildings Conference and
  Exhibition}}.
\newblock


\bibitem[\protect\citeauthoryear{Huizenga, Abbaszadeh, Zagreus, and
  Arens}{Huizenga et~al\mbox{.}}{2006}]%
        {huizenga2006air}
\bibfield{author}{\bibinfo{person}{Charlie Huizenga}, \bibinfo{person}{Sahar
  Abbaszadeh}, \bibinfo{person}{Leah Zagreus}, {and} \bibinfo{person}{Edward~A
  Arens}.} \bibinfo{year}{2006}\natexlab{}.
\newblock \showarticletitle{Air Quality and Thermal Comfort in Office
  Buildings: Results of a Large Indoor Environmental Quality Survey}.
\newblock \bibinfo{journal}{\emph{Proceeding of Healthy Buildings 2006}}
  \bibinfo{volume}{3} (\bibinfo{year}{2006}), \bibinfo{pages}{393--397}.
\newblock


\bibitem[\protect\citeauthoryear{Iaffaldano and Muchinsky}{Iaffaldano and
  Muchinsky}{1985}]%
        {iaffaldano1985job}
\bibfield{author}{\bibinfo{person}{Michelle~T Iaffaldano} {and}
  \bibinfo{person}{Paul~M Muchinsky}.} \bibinfo{year}{1985}\natexlab{}.
\newblock \showarticletitle{Job Satisfaction and Job Performance: A
  Meta-analysis.}
\newblock \bibinfo{journal}{\emph{Psychological bulletin}}
  \bibinfo{volume}{97}, \bibinfo{number}{2} (\bibinfo{year}{1985}),
  \bibinfo{pages}{251}.
\newblock


\bibitem[\protect\citeauthoryear{Ishii, Kanagawa, Shimamura, Uchiyama, Miyagi,
  Obayashi, and Shimoda}{Ishii et~al\mbox{.}}{2018}]%
        {ishii2018intellectual}
\bibfield{author}{\bibinfo{person}{Hirotake Ishii}, \bibinfo{person}{Hidehiro
  Kanagawa}, \bibinfo{person}{Yuta Shimamura}, \bibinfo{person}{Kosuke
  Uchiyama}, \bibinfo{person}{Kazune Miyagi}, \bibinfo{person}{Fumiaki
  Obayashi}, {and} \bibinfo{person}{Hiroshi Shimoda}.}
  \bibinfo{year}{2018}\natexlab{}.
\newblock \showarticletitle{Intellectual productivity under task ambient
  lighting}.
\newblock \bibinfo{journal}{\emph{Lighting Research \& Technology}}
  \bibinfo{volume}{50}, \bibinfo{number}{2} (\bibinfo{year}{2018}),
  \bibinfo{pages}{237--252}.
\newblock


\bibitem[\protect\citeauthoryear{James, Demaree, and Wolf}{James
  et~al\mbox{.}}{1984}]%
        {james1984estimating}
\bibfield{author}{\bibinfo{person}{Lawrence~R James}, \bibinfo{person}{Robert~G
  Demaree}, {and} \bibinfo{person}{Gerrit Wolf}.}
  \bibinfo{year}{1984}\natexlab{}.
\newblock \showarticletitle{Estimating within-group interrater reliability with
  and without response bias.}
\newblock \bibinfo{journal}{\emph{Journal of applied psychology}}
  \bibinfo{volume}{69}, \bibinfo{number}{1} (\bibinfo{year}{1984}),
  \bibinfo{pages}{85}.
\newblock


\bibitem[\protect\citeauthoryear{Jin, Bekiaris-Liberis, Weekly, Spanos, and
  Bayen}{Jin et~al\mbox{.}}{2016}]%
        {jin2016occupancy}
\bibfield{author}{\bibinfo{person}{Ming Jin}, \bibinfo{person}{Nikolaos
  Bekiaris-Liberis}, \bibinfo{person}{Kevin Weekly}, \bibinfo{person}{Costas~J
  Spanos}, {and} \bibinfo{person}{Alexandre~M Bayen}.}
  \bibinfo{year}{2016}\natexlab{}.
\newblock \showarticletitle{Occupancy Detection via Environmental Sensing}.
\newblock \bibinfo{journal}{\emph{IEEE Transactions on Automation Science and
  Engineering}} \bibinfo{volume}{15}, \bibinfo{number}{2}
  (\bibinfo{year}{2016}), \bibinfo{pages}{443--455}.
\newblock


\bibitem[\protect\citeauthoryear{Jin, Zou, Weekly, Jia, Bayen, and Spanos}{Jin
  et~al\mbox{.}}{2014}]%
        {jin2014environmental}
\bibfield{author}{\bibinfo{person}{Ming Jin}, \bibinfo{person}{Han Zou},
  \bibinfo{person}{Kevin Weekly}, \bibinfo{person}{Ruoxi Jia},
  \bibinfo{person}{Alexandre~M Bayen}, {and} \bibinfo{person}{Costas~J
  Spanos}.} \bibinfo{year}{2014}\natexlab{}.
\newblock \showarticletitle{Environmental Sensing by Wearable Device for Indoor
  Activity and Location Estimation}. In \bibinfo{booktitle}{\emph{IECON
  2014-40th Annual Conference of the IEEE Industrial Electronics Society}}.
  IEEE, \bibinfo{pages}{5369--5375}.
\newblock


\bibitem[\protect\citeauthoryear{Kaiser}{Kaiser}{1960}]%
        {kaiser1960application}
\bibfield{author}{\bibinfo{person}{Henry~F Kaiser}.}
  \bibinfo{year}{1960}\natexlab{}.
\newblock \showarticletitle{The Application of Electronic Computers to Factor
  Analysis}.
\newblock \bibinfo{journal}{\emph{Educational and Psychological Measurement}}
  \bibinfo{volume}{20}, \bibinfo{number}{1} (\bibinfo{year}{1960}),
  \bibinfo{pages}{141--151}.
\newblock


\bibitem[\protect\citeauthoryear{Kang, Ou, and Mak}{Kang et~al\mbox{.}}{2017}]%
        {kang2017impact}
\bibfield{author}{\bibinfo{person}{Shengxian Kang}, \bibinfo{person}{Dayi Ou},
  {and} \bibinfo{person}{Cheuk~Ming Mak}.} \bibinfo{year}{2017}\natexlab{}.
\newblock \showarticletitle{The Impact of Indoor Environmental Quality on Work
  Productivity in University Open-plan Research Offices}.
\newblock \bibinfo{journal}{\emph{Building and Environment}}
  \bibinfo{volume}{124} (\bibinfo{year}{2017}), \bibinfo{pages}{78--89}.
\newblock


\bibitem[\protect\citeauthoryear{Karami, McMorrow, and Wang}{Karami
  et~al\mbox{.}}{2018}]%
        {karami2018continuous}
\bibfield{author}{\bibinfo{person}{Majid Karami},
  \bibinfo{person}{Gabrielle~Viola McMorrow}, {and} \bibinfo{person}{Liping
  Wang}.} \bibinfo{year}{2018}\natexlab{}.
\newblock \showarticletitle{Continuous monitoring of indoor environmental
  quality using an Arduino-based data acquisition system}.
\newblock \bibinfo{journal}{\emph{Journal of Building Engineering}}
  \bibinfo{volume}{19} (\bibinfo{year}{2018}), \bibinfo{pages}{412--419}.
\newblock


\bibitem[\protect\citeauthoryear{Kauffeld and Lehmann-Willenbrock}{Kauffeld and
  Lehmann-Willenbrock}{2012}]%
        {kauffeld2012meetings}
\bibfield{author}{\bibinfo{person}{Simone Kauffeld} {and} \bibinfo{person}{Nale
  Lehmann-Willenbrock}.} \bibinfo{year}{2012}\natexlab{}.
\newblock \showarticletitle{Meetings Matter: Effects of Team Meetings on Team
  and Organizational Success}.
\newblock \bibinfo{journal}{\emph{Small Group Research}} \bibinfo{volume}{43},
  \bibinfo{number}{2} (\bibinfo{year}{2012}), \bibinfo{pages}{130--158}.
\newblock


\bibitem[\protect\citeauthoryear{Kim}{Kim}{2019}]%
        {meeting_lifesize}
\bibfield{author}{\bibinfo{person}{Jasmine Kim}.}
  \bibinfo{year}{2019}\natexlab{}.
\newblock \bibinfo{booktitle}{\emph{6 Most Common Types of Business Meetings}}.
\newblock
\urldef\tempurl%
\url{https://www.lifesize.com/en/video-conferencing-blog/types-of-business-meetings}
\showURL{%
\tempurl}


\bibitem[\protect\citeauthoryear{Kim, de~Dear, Candido, Zhang, and Arens}{Kim
  et~al\mbox{.}}{2013}]%
        {kim2013gender}
\bibfield{author}{\bibinfo{person}{Jungsoo Kim}, \bibinfo{person}{Richard de
  Dear}, \bibinfo{person}{Christhina Candido}, \bibinfo{person}{Hui Zhang},
  {and} \bibinfo{person}{Edward Arens}.} \bibinfo{year}{2013}\natexlab{}.
\newblock \showarticletitle{Gender Differences in Office Occupant Perception of
  Indoor Environmental Quality (IEQ)}.
\newblock \bibinfo{journal}{\emph{Building and Environment}}
  \bibinfo{volume}{70} (\bibinfo{year}{2013}), \bibinfo{pages}{245--256}.
\newblock


\bibitem[\protect\citeauthoryear{Lee, Lam, and Fai}{Lee et~al\mbox{.}}{2001}]%
        {lee2001characterization}
\bibfield{author}{\bibinfo{person}{SC Lee}, \bibinfo{person}{Sanches Lam},
  {and} \bibinfo{person}{Ho~Kin Fai}.} \bibinfo{year}{2001}\natexlab{}.
\newblock \showarticletitle{Characterization of VOCs, Ozone, and PM10 Emissions
  from Office Equipment in an Environmental Chamber}.
\newblock \bibinfo{journal}{\emph{Building and Environment}}
  \bibinfo{volume}{36}, \bibinfo{number}{7} (\bibinfo{year}{2001}),
  \bibinfo{pages}{837--842}.
\newblock


\bibitem[\protect\citeauthoryear{Lent}{Lent}{2015}]%
        {lent2015_meetings}
\bibfield{author}{\bibinfo{person}{Richard~M Lent}.}
  \bibinfo{year}{2015}\natexlab{}.
\newblock \bibinfo{booktitle}{\emph{Leading Great Meetings: How to Structure
  Yours for Success}}.
\newblock \bibinfo{publisher}{Meeting for Results}.
\newblock


\bibitem[\protect\citeauthoryear{Leslie A.~Perlow}{Leslie A.~Perlow}{2017}]%
        {hbr_meeting_madness}
\bibfield{author}{\bibinfo{person}{Eunice~Eun Leslie A.~Perlow, Constance
  Noonan~Hadley}.} \bibinfo{year}{2017}\natexlab{}.
\newblock \bibinfo{booktitle}{\emph{Stop the Meeting Madness}}.
\newblock
\urldef\tempurl%
\url{https://hbr.org/2017/07/stop-the-meeting-madness}
\showURL{%
\tempurl}


\bibitem[\protect\citeauthoryear{Lloret, Ferreres, Hern{\'a}ndez, and
  Tom{\'a}s}{Lloret et~al\mbox{.}}{2017}]%
        {lloret2017exploratory}
\bibfield{author}{\bibinfo{person}{Susana Lloret},
  \bibinfo{person}{Adoraci{\'o}n Ferreres}, \bibinfo{person}{Ana
  Hern{\'a}ndez}, {and} \bibinfo{person}{In{\'e}s Tom{\'a}s}.}
  \bibinfo{year}{2017}\natexlab{}.
\newblock \showarticletitle{The Exploratory Factor Analysis of Items: Guided
  Analysis Based on Empirical Data and Software}.
\newblock \bibinfo{journal}{\emph{Anales de Psicolog{\'\i}a}}
  \bibinfo{volume}{33}, \bibinfo{number}{2} (\bibinfo{year}{2017}),
  \bibinfo{pages}{417--432}.
\newblock


\bibitem[\protect\citeauthoryear{MacNaughton, Pegues, Satish, Santanam,
  Spengler, and Allen}{MacNaughton et~al\mbox{.}}{2015}]%
        {macnaughton2015economic}
\bibfield{author}{\bibinfo{person}{Piers MacNaughton}, \bibinfo{person}{James
  Pegues}, \bibinfo{person}{Usha Satish}, \bibinfo{person}{Suresh Santanam},
  \bibinfo{person}{John Spengler}, {and} \bibinfo{person}{Joseph Allen}.}
  \bibinfo{year}{2015}\natexlab{}.
\newblock \showarticletitle{Economic, Environmental and Health Implications of
  Enhanced Ventilation in Office Buildings}.
\newblock \bibinfo{journal}{\emph{International Journal of Environmental
  Research and Public Health}} \bibinfo{volume}{12}, \bibinfo{number}{11}
  (\bibinfo{year}{2015}), \bibinfo{pages}{14709--14722}.
\newblock


\bibitem[\protect\citeauthoryear{MeetingSift}{MeetingSift}{2020}]%
        {meeting_sift}
\bibfield{author}{\bibinfo{person}{MeetingSift}.}
  \bibinfo{year}{2020}\natexlab{}.
\newblock \bibinfo{booktitle}{\emph{The Six Most Common Types of Meetings}}.
\newblock
\urldef\tempurl%
\url{http://meetingsift.com/the-six-types-of-meetings/}
\showURL{%
\tempurl}


\bibitem[\protect\citeauthoryear{Meister}{Meister}{2018}]%
        {hbr_office_light}
\bibfield{author}{\bibinfo{person}{Jeanne~C. Meister}.}
  \bibinfo{year}{2018}\natexlab{}.
\newblock \showarticletitle{The \#1 Office Perk? Natural Light}.
\newblock \bibinfo{journal}{\emph{Harvard Business Review}}
  (\bibinfo{year}{2018}).
\newblock


\bibitem[\protect\citeauthoryear{Mendell, Naco, Wilcox, and Sieber}{Mendell
  et~al\mbox{.}}{2003}]%
        {mendell2003environmental}
\bibfield{author}{\bibinfo{person}{Mark~J Mendell}, \bibinfo{person}{Gina~M
  Naco}, \bibinfo{person}{Thomas~G Wilcox}, {and} \bibinfo{person}{W~Karl
  Sieber}.} \bibinfo{year}{2003}\natexlab{}.
\newblock \showarticletitle{Environmental Risk Factors and Work-related Lower
  Respiratory Symptoms in 80 Office Buildings: An Exploratory Analysis of NIOSH
  data}.
\newblock \bibinfo{journal}{\emph{American Journal of Industrial Medicine}}
  \bibinfo{volume}{43}, \bibinfo{number}{6} (\bibinfo{year}{2003}),
  \bibinfo{pages}{630--641}.
\newblock


\bibitem[\protect\citeauthoryear{Minute}{Minute}{2020}]%
        {meeting_minute}
\bibfield{author}{\bibinfo{person}{Minute}.} \bibinfo{year}{2020}\natexlab{}.
\newblock \bibinfo{booktitle}{\emph{5 Types of Meetings: All You Need to Know
  for Successful Meetings}}.
\newblock
\urldef\tempurl%
\url{https://www.getminute.com/types-of-meetings/}
\showURL{%
\tempurl}


\bibitem[\protect\citeauthoryear{Mirjafari, Masaba, Grover, Wang, Audia,
  Campbell, Chawla, Swain, Choudhury, Dey, et~al\mbox{.}}{Mirjafari
  et~al\mbox{.}}{2019}]%
        {mirjafari2019differentiating}
\bibfield{author}{\bibinfo{person}{Shayan Mirjafari}, \bibinfo{person}{Kizito
  Masaba}, \bibinfo{person}{Ted Grover}, \bibinfo{person}{Weichen Wang},
  \bibinfo{person}{Pino Audia}, \bibinfo{person}{Andrew~T Campbell},
  \bibinfo{person}{Nitesh~V Chawla}, \bibinfo{person}{Vedant~Das Swain},
  \bibinfo{person}{Munmun~De Choudhury}, \bibinfo{person}{Anind~K Dey},
  {et~al\mbox{.}}} \bibinfo{year}{2019}\natexlab{}.
\newblock \showarticletitle{Differentiating Higher and Lower Job Performers in
  the Workplace Using Mobile Sensing}.
\newblock \bibinfo{journal}{\emph{Proceedings of the ACM on Interactive,
  Mobile, Wearable and Ubiquitous Technologies}} \bibinfo{volume}{3},
  \bibinfo{number}{2} (\bibinfo{year}{2019}), \bibinfo{pages}{1--24}.
\newblock


\bibitem[\protect\citeauthoryear{Mitchell, Zhang, Sigsgaard, Jantunen, Lioy,
  Samson, and Karol}{Mitchell et~al\mbox{.}}{2007}]%
        {mitchell2007current}
\bibfield{author}{\bibinfo{person}{Clifford~S Mitchell},
  \bibinfo{person}{Junfeng Zhang}, \bibinfo{person}{Torben Sigsgaard},
  \bibinfo{person}{Matti Jantunen}, \bibinfo{person}{Paul~J Lioy},
  \bibinfo{person}{Robert Samson}, {and} \bibinfo{person}{Meryl~H Karol}.}
  \bibinfo{year}{2007}\natexlab{}.
\newblock \showarticletitle{Current State of the Science: Health Effects and
  Indoor Environmental Quality}.
\newblock \bibinfo{journal}{\emph{Environmental Health Perspectives}}
  \bibinfo{volume}{115}, \bibinfo{number}{6} (\bibinfo{year}{2007}),
  \bibinfo{pages}{958--964}.
\newblock


\bibitem[\protect\citeauthoryear{Mui and Wong}{Mui and Wong}{2006}]%
        {mui2006acceptable}
\bibfield{author}{\bibinfo{person}{KW Mui} {and} \bibinfo{person}{LT Wong}.}
  \bibinfo{year}{2006}\natexlab{}.
\newblock \showarticletitle{Acceptable Illumination Levels for Office
  occupants}.
\newblock \bibinfo{journal}{\emph{Architectural Science Review}}
  \bibinfo{volume}{49}, \bibinfo{number}{2} (\bibinfo{year}{2006}),
  \bibinfo{pages}{116--119}.
\newblock


\bibitem[\protect\citeauthoryear{Mujan, An{\dj}elkovi{\'c}, Mun{\'c}an,
  Kljaji{\'c}, and Ru{\v{z}}i{\'c}}{Mujan et~al\mbox{.}}{2019}]%
        {mujan2019influence}
\bibfield{author}{\bibinfo{person}{Igor Mujan}, \bibinfo{person}{Aleksandar~S
  An{\dj}elkovi{\'c}}, \bibinfo{person}{Vladimir Mun{\'c}an},
  \bibinfo{person}{Miroslav Kljaji{\'c}}, {and} \bibinfo{person}{Dragan
  Ru{\v{z}}i{\'c}}.} \bibinfo{year}{2019}\natexlab{}.
\newblock \showarticletitle{Influence of indoor environmental quality on human
  health and productivity-A review}.
\newblock \bibinfo{journal}{\emph{Journal of cleaner production}}
  \bibinfo{volume}{217} (\bibinfo{year}{2019}), \bibinfo{pages}{646--657}.
\newblock


\bibitem[\protect\citeauthoryear{Neidell}{Neidell}{2017}]%
        {neidell2017air}
\bibfield{author}{\bibinfo{person}{Matthew Neidell}.}
  \bibinfo{year}{2017}\natexlab{}.
\newblock \showarticletitle{Air pollution and worker productivity}.
\newblock \bibinfo{journal}{\emph{IZA World of Labor}} (\bibinfo{year}{2017}).
\newblock


\bibitem[\protect\citeauthoryear{Niemantsverdriet and
  Erickson}{Niemantsverdriet and Erickson}{2017}]%
        {niemantsverdriet2017recurring}
\bibfield{author}{\bibinfo{person}{Karin Niemantsverdriet} {and}
  \bibinfo{person}{Thomas Erickson}.} \bibinfo{year}{2017}\natexlab{}.
\newblock \showarticletitle{Recurring Meetings: An Experiential Account of
  Repeating Meetings in a Large Organization}.
\newblock \bibinfo{journal}{\emph{Proceedings of the ACM on Human-Computer
  Interaction}} \bibinfo{volume}{1}, \bibinfo{number}{CSCW}
  (\bibinfo{year}{2017}), \bibinfo{pages}{84}.
\newblock


\bibitem[\protect\citeauthoryear{Niemel{\"a}, Hannula, Rautio, Reijula, and
  Railio}{Niemel{\"a} et~al\mbox{.}}{2002}]%
        {niemela2002effect}
\bibfield{author}{\bibinfo{person}{Raimo Niemel{\"a}}, \bibinfo{person}{Mika
  Hannula}, \bibinfo{person}{Sari Rautio}, \bibinfo{person}{Kari Reijula},
  {and} \bibinfo{person}{Jorma Railio}.} \bibinfo{year}{2002}\natexlab{}.
\newblock \showarticletitle{The effect of air temperature on labour
  productivity in call centres—a case study}.
\newblock \bibinfo{journal}{\emph{Energy and Buildings}} \bibinfo{volume}{34},
  \bibinfo{number}{8} (\bibinfo{year}{2002}), \bibinfo{pages}{759--764}.
\newblock


\bibitem[\protect\citeauthoryear{Norman and Streiner}{Norman and
  Streiner}{2008}]%
        {norman2008biostatistics}
\bibfield{author}{\bibinfo{person}{Geoffrey~R Norman} {and}
  \bibinfo{person}{David~L Streiner}.} \bibinfo{year}{2008}\natexlab{}.
\newblock \bibinfo{booktitle}{\emph{Biostatistics: The Bare Essentials}}.
\newblock \bibinfo{publisher}{PMPH}.
\newblock


\bibitem[\protect\citeauthoryear{Nunally and Bernstein}{Nunally and
  Bernstein}{1978}]%
        {nunally1978psychometric}
\bibfield{author}{\bibinfo{person}{Jum~C Nunally} {and} \bibinfo{person}{Ira~H
  Bernstein}.} \bibinfo{year}{1978}\natexlab{}.
\newblock \bibinfo{title}{Psychometric Theory}.
\newblock
\newblock


\bibitem[\protect\citeauthoryear{O'Neill}{O'Neill}{2017}]%
        {o2017overview}
\bibfield{author}{\bibinfo{person}{Thomas~A O'Neill}.}
  \bibinfo{year}{2017}\natexlab{}.
\newblock \showarticletitle{An overview of interrater agreement on Likert
  scales for researchers and practitioners}.
\newblock \bibinfo{journal}{\emph{Frontiers in psychology}}
  \bibinfo{volume}{8} (\bibinfo{year}{2017}), \bibinfo{pages}{777}.
\newblock


\bibitem[\protect\citeauthoryear{Redlich, Sparer, and Cullen}{Redlich
  et~al\mbox{.}}{1997}]%
        {redlich1997sick}
\bibfield{author}{\bibinfo{person}{Carrie~A Redlich}, \bibinfo{person}{Judy
  Sparer}, {and} \bibinfo{person}{Mark~R Cullen}.}
  \bibinfo{year}{1997}\natexlab{}.
\newblock \showarticletitle{Sick-Building Syndrome}.
\newblock \bibinfo{journal}{\emph{The Lancet}} \bibinfo{volume}{349},
  \bibinfo{number}{9057} (\bibinfo{year}{1997}), \bibinfo{pages}{1013--1016}.
\newblock


\bibitem[\protect\citeauthoryear{Rogelberg, Shanock, and Scott}{Rogelberg
  et~al\mbox{.}}{2012}]%
        {rogelberg2012wasted}
\bibfield{author}{\bibinfo{person}{Steven~G Rogelberg},
  \bibinfo{person}{Linda~Rhoades Shanock}, {and} \bibinfo{person}{Cliff~W
  Scott}.} \bibinfo{year}{2012}\natexlab{}.
\newblock \showarticletitle{Wasted Time and Money in Meetings: Increasing
  Return on Investment}.
\newblock \bibinfo{journal}{\emph{Small Group Research}} \bibinfo{volume}{43},
  \bibinfo{number}{2} (\bibinfo{year}{2012}), \bibinfo{pages}{236--245}.
\newblock


\bibitem[\protect\citeauthoryear{Rolstad, Adler, and Ryd{\'e}n}{Rolstad
  et~al\mbox{.}}{2011}]%
        {rolstad2011response}
\bibfield{author}{\bibinfo{person}{Sindre Rolstad}, \bibinfo{person}{John
  Adler}, {and} \bibinfo{person}{Anna Ryd{\'e}n}.}
  \bibinfo{year}{2011}\natexlab{}.
\newblock \showarticletitle{Response Burden and Questionnaire Length: Is
  Shorter Better? A Review and Meta-Analysis}.
\newblock \bibinfo{journal}{\emph{Value in Health}} \bibinfo{volume}{14},
  \bibinfo{number}{8} (\bibinfo{year}{2011}), \bibinfo{pages}{1101--1108}.
\newblock


\bibitem[\protect\citeauthoryear{Romano and Nunamaker}{Romano and
  Nunamaker}{2001}]%
        {romano2001meeting}
\bibfield{author}{\bibinfo{person}{Nicholas~C Romano} {and}
  \bibinfo{person}{Jay~F Nunamaker}.} \bibinfo{year}{2001}\natexlab{}.
\newblock \showarticletitle{Meeting Analysis: Findings from Research and
  Practice}. In \bibinfo{booktitle}{\emph{Proceedings of the 34th Annual Hawaii
  International Conference on System Sciences}}. IEEE, \bibinfo{pages}{13--pp}.
\newblock


\bibitem[\protect\citeauthoryear{Rothmann and Coetzer}{Rothmann and
  Coetzer}{2003}]%
        {rothmann2003big}
\bibfield{author}{\bibinfo{person}{Sebastiaan Rothmann} {and}
  \bibinfo{person}{Elize~P Coetzer}.} \bibinfo{year}{2003}\natexlab{}.
\newblock \showarticletitle{The Big Five Personality Dimensions and Job
  Performance}.
\newblock \bibinfo{journal}{\emph{SA Journal of Industrial Psychology}}
  \bibinfo{volume}{29}, \bibinfo{number}{1} (\bibinfo{year}{2003}),
  \bibinfo{pages}{68--74}.
\newblock


\bibitem[\protect\citeauthoryear{Russel}{Russel}{2015}]%
        {attentiv}
\bibfield{author}{\bibinfo{person}{Daniel Russel}.}
  \bibinfo{year}{2015}\natexlab{}.
\newblock \bibinfo{booktitle}{\emph{America Meets a lot. An Analysis of Meeting
  Length, Frequency and Cost}}.
\newblock
\urldef\tempurl%
\url{http://attentiv.com/america-meets-a-lot/}
\showURL{%
\tempurl}


\bibitem[\protect\citeauthoryear{Sartori, Costantini, Ceschi, and
  Scalco}{Sartori et~al\mbox{.}}{2017}]%
        {sartori2017not}
\bibfield{author}{\bibinfo{person}{Riccardo Sartori}, \bibinfo{person}{Arianna
  Costantini}, \bibinfo{person}{Andrea Ceschi}, {and} \bibinfo{person}{Andrea
  Scalco}.} \bibinfo{year}{2017}\natexlab{}.
\newblock \showarticletitle{Not Only Correlations: A Different Approach for
  Investigating the Relationship Between the Big Five Personality Traits and
  Job Performance Based on Workers and Employees’ Perception}.
\newblock \bibinfo{journal}{\emph{Quality \& Quantity}} \bibinfo{volume}{51},
  \bibinfo{number}{6} (\bibinfo{year}{2017}), \bibinfo{pages}{2507--2519}.
\newblock


\bibitem[\protect\citeauthoryear{Sbar, Podbelski, Yang, and Pease}{Sbar
  et~al\mbox{.}}{2012}]%
        {sbar2012electrochromic}
\bibfield{author}{\bibinfo{person}{Neil~L Sbar}, \bibinfo{person}{Lou
  Podbelski}, \bibinfo{person}{Hong~Mo Yang}, {and} \bibinfo{person}{Brad
  Pease}.} \bibinfo{year}{2012}\natexlab{}.
\newblock \showarticletitle{Electrochromic Dynamic Windows for Office
  Buildings}.
\newblock \bibinfo{journal}{\emph{International Journal of Sustainable Built
  Environment}} \bibinfo{volume}{1}, \bibinfo{number}{1}
  (\bibinfo{year}{2012}), \bibinfo{pages}{125--139}.
\newblock


\bibitem[\protect\citeauthoryear{Schaule, Johanssen, Bruegge, and
  Loftness}{Schaule et~al\mbox{.}}{2018}]%
        {schaule2018employing}
\bibfield{author}{\bibinfo{person}{Florian Schaule}, \bibinfo{person}{Jan~Ole
  Johanssen}, \bibinfo{person}{Bernd Bruegge}, {and} \bibinfo{person}{Vivian
  Loftness}.} \bibinfo{year}{2018}\natexlab{}.
\newblock \showarticletitle{Employing Consumer Wearables to Detect Office
  Workers' Cognitive Load for Interruption Management}.
\newblock \bibinfo{journal}{\emph{Proceedings of the ACM on Interactive,
  Mobile, Wearable and Ubiquitous Technologies}} \bibinfo{volume}{2},
  \bibinfo{number}{1} (\bibinfo{year}{2018}), \bibinfo{pages}{1--20}.
\newblock


\bibitem[\protect\citeauthoryear{Schwarz}{Schwarz}{2016}]%
        {hbr_schwarz_start}
\bibfield{author}{\bibinfo{person}{Roger Schwarz}.}
  \bibinfo{year}{2016}\natexlab{}.
\newblock \showarticletitle{8 Ground Rules for Great Meetings}.
\newblock \bibinfo{journal}{\emph{Harvard Business Review}}
  (\bibinfo{year}{2016}).
\newblock


\bibitem[\protect\citeauthoryear{Sepp{\"a}nen and Fisk}{Sepp{\"a}nen and
  Fisk}{2006}]%
        {seppanen2006some}
\bibfield{author}{\bibinfo{person}{Olli~A Sepp{\"a}nen} {and}
  \bibinfo{person}{William Fisk}.} \bibinfo{year}{2006}\natexlab{}.
\newblock \showarticletitle{Some Quantitative Relations Between Indoor
  Environmental Quality and Work Performance or Health}.
\newblock \bibinfo{journal}{\emph{Hvac\&R Research}} \bibinfo{volume}{12},
  \bibinfo{number}{4} (\bibinfo{year}{2006}), \bibinfo{pages}{957--973}.
\newblock


\bibitem[\protect\citeauthoryear{Snow}{Snow}{2018}]%
        {snow2018indoor}
\bibfield{author}{\bibinfo{person}{Stephen Snow}.}
  \bibinfo{year}{2018}\natexlab{}.
\newblock \showarticletitle{Indoor Air Quality: Opportunities for Behaviour
  Change Towards Healthier Offices, a Two-part Report}.
\newblock  (\bibinfo{year}{2018}).
\newblock


\bibitem[\protect\citeauthoryear{Song, Mao, and Liu}{Song
  et~al\mbox{.}}{2019}]%
        {song2019human}
\bibfield{author}{\bibinfo{person}{Ying Song}, \bibinfo{person}{Fubing Mao},
  {and} \bibinfo{person}{Qing Liu}.} \bibinfo{year}{2019}\natexlab{}.
\newblock \showarticletitle{Human Comfort in Indoor Environment: A Review on
  Assessment Criteria, Data Collection and Data Analysis Methods}.
\newblock \bibinfo{journal}{\emph{IEEE Access}}  \bibinfo{volume}{7}
  (\bibinfo{year}{2019}), \bibinfo{pages}{119774--119786}.
\newblock


\bibitem[\protect\citeauthoryear{Sundell}{Sundell}{2004}]%
        {sundell2004history}
\bibfield{author}{\bibinfo{person}{Jan Sundell}.}
  \bibinfo{year}{2004}\natexlab{}.
\newblock \showarticletitle{On the History of Indoor Air Quality and Health}.
\newblock \bibinfo{journal}{\emph{Indoor Air}} \bibinfo{volume}{14},
  \bibinfo{number}{s 7} (\bibinfo{year}{2004}), \bibinfo{pages}{51--58}.
\newblock


\bibitem[\protect\citeauthoryear{Tom Y.~Chang and Neidell}{Tom Y.~Chang and
  Neidell}{2016}]%
        {hbr_air_pollution}
\bibfield{author}{\bibinfo{person}{Tal~Gross Tom Y.~Chang, Joshua Graff~Zivin}
  {and} \bibinfo{person}{Matthew Neidell}.} \bibinfo{year}{2016}\natexlab{}.
\newblock \showarticletitle{Air Pollution Is Making Office Workers Less
  Productive}.
\newblock \bibinfo{journal}{\emph{Harvard Business Review}}
  (\bibinfo{year}{2016}).
\newblock


\bibitem[\protect\citeauthoryear{Viswesvaran and Ones}{Viswesvaran and
  Ones}{2000}]%
        {viswesvaran2000perspectives}
\bibfield{author}{\bibinfo{person}{Chockalingam Viswesvaran} {and}
  \bibinfo{person}{Deniz~S Ones}.} \bibinfo{year}{2000}\natexlab{}.
\newblock \showarticletitle{Perspectives on Models of Job Performance}.
\newblock \bibinfo{journal}{\emph{International Journal of Selection and
  Assessment}} \bibinfo{volume}{8}, \bibinfo{number}{4} (\bibinfo{year}{2000}),
  \bibinfo{pages}{216--226}.
\newblock


\bibitem[\protect\citeauthoryear{Wang, Wan, Li, and Zhang}{Wang
  et~al\mbox{.}}{2016}]%
        {wang2016implementing}
\bibfield{author}{\bibinfo{person}{Shiyong Wang}, \bibinfo{person}{Jiafu Wan},
  \bibinfo{person}{Di Li}, {and} \bibinfo{person}{Chunhua Zhang}.}
  \bibinfo{year}{2016}\natexlab{}.
\newblock \showarticletitle{Implementing Smart Factory of Industrie 4.0: An
  Outlook}.
\newblock \bibinfo{journal}{\emph{International journal of distributed sensor
  networks}} \bibinfo{volume}{12}, \bibinfo{number}{1} (\bibinfo{year}{2016}),
  \bibinfo{pages}{3159805}.
\newblock


\bibitem[\protect\citeauthoryear{Wargocki, Sundell, Bischof, Brundrett, Fanger,
  Gyntelberg, Hanssen, Harrison, Pickering, Sepp{\"a}nen,
  et~al\mbox{.}}{Wargocki et~al\mbox{.}}{2002}]%
        {wargocki2002ventilation}
\bibfield{author}{\bibinfo{person}{Pawel Wargocki}, \bibinfo{person}{Jan
  Sundell}, \bibinfo{person}{W Bischof}, \bibinfo{person}{G Brundrett},
  \bibinfo{person}{Povl~Ole Fanger}, \bibinfo{person}{F Gyntelberg},
  \bibinfo{person}{SO Hanssen}, \bibinfo{person}{P Harrison},
  \bibinfo{person}{A Pickering}, \bibinfo{person}{O Sepp{\"a}nen},
  {et~al\mbox{.}}} \bibinfo{year}{2002}\natexlab{}.
\newblock \showarticletitle{Ventilation and health in non-industrial indoor
  environments: report from a European Multidisciplinary Scientific Consensus
  Meeting (EUROVEN)}.
\newblock \bibinfo{journal}{\emph{Indoor air}} \bibinfo{volume}{12},
  \bibinfo{number}{2} (\bibinfo{year}{2002}), \bibinfo{pages}{113--128}.
\newblock


\bibitem[\protect\citeauthoryear{Wyon}{Wyon}{2004}]%
        {wyon2004effects}
\bibfield{author}{\bibinfo{person}{David~P Wyon}.}
  \bibinfo{year}{2004}\natexlab{}.
\newblock \showarticletitle{The Effects of Indoor Air Quality on Performance
  and Productivity}.
\newblock \bibinfo{journal}{\emph{Indoor Air}}  \bibinfo{volume}{14}
  (\bibinfo{year}{2004}), \bibinfo{pages}{92--101}.
\newblock


\bibitem[\protect\citeauthoryear{Zhong, Alavi, and Lalanne}{Zhong
  et~al\mbox{.}}{2020}]%
        {zhong2020hilo}
\bibfield{author}{\bibinfo{person}{Sailin Zhong}, \bibinfo{person}{Hamed~S
  Alavi}, {and} \bibinfo{person}{Denis Lalanne}.}
  \bibinfo{year}{2020}\natexlab{}.
\newblock \showarticletitle{Hilo-wear: Exploring Wearable Interaction with
  Indoor Air Quality Forecast}. In \bibinfo{booktitle}{\emph{Extended Abstracts
  of the 2020 CHI Conference on Human Factors in Computing Systems Extended
  Abstracts}}. \bibinfo{pages}{1--8}.
\newblock


\end{thebibliography}

\end{document}